\documentclass[prd,preprint,superscriptaddress,showpacs,byrevtex]{revtex4}
\usepackage{bm}
\def\fsl#1{\setbox0=\hbox{$#1$}           
   \dimen0=\wd0                                 
   \setbox1=\hbox{/} \dimen1=\wd1               
   \ifdim\dimen0>\dimen1                        
      \rlap{\hbox to \dimen0{\hfil/\hfil}}      
      #1                                        
   \else                                        
      \rlap{\hbox to \dimen1{\hfil$#1$\hfil}}   
      /                                         
   \fi}                                         %
\newcommand{\be}{\begin{equation}}
\newcommand{\ee}{\end{equation}}
\newcommand{\bea}{\begin{eqnarray}}
\newcommand{\eea}{\end{eqnarray}}
\newcommand{\beq}{\begin{equation}}
\newcommand{\eeq}{\end{equation}}
\newcommand{\beqs}{\begin{eqnarray}}
\newcommand{\eeqs}{\end{eqnarray}}

\begin{document}
\title{ Proof of NRQCD Factorization at All Order in Coupling Constant in Heavy Quarkonium Production }
\author{Gouranga C Nayak }\email{nayak@max2.physics.sunysb.edu}
\affiliation{ 665 East Pine Street, Long Beach, New York 11561, USA }
\date{\today}
\begin{abstract}
Recently the proof of factorization in heavy quarkonium production in NRQCD color octet mechanism is given
at next-to-next-to-leading order (NNLO) in coupling constant by using diagrammatic method of QCD. In this paper
we prove factorization in heavy quarkonium production in NRQCD color octet mechanism at all order in coupling constant
by using path integral method of QCD. Our proof
is valid to all powers in the heavy quark relative velocity. We find that the gauge invariance and the factorization at all
order in coupling constant require gauge-completed non-perturbative NRQCD
matrix elements that were introduced previously to prove factorization at NNLO.
\end{abstract}
\maketitle
\pagestyle{plain}
\pagenumbering{arabic}
\section{Introduction}
In the last two decades, the NRQCD color octet mechanism \cite{nrqcd} for heavy quarkonium production has been very successful
in explaining experimental data at high energy colliders such as at Tevatron \cite{tevatron}
and at LHC \cite{lhc}. In its original formulation \cite{nrqcd} the proof of factorization
in heavy quarkonium production in NRQCD color octet mechanism was lacking.
The proof of factorization is an essential requirement
to study heavy quarkonium production at high energy colliders.
Factorization refers to separation of short-distance effects from long-distance effects in quantum field theory.

Recently the proof of factorization
in heavy quarkonium production in NRQCD color octet mechanism is given at next-to-next to leading order (NLLO) in coupling constant
by using diagrammatic method of QCD \cite{nyka,nykb,nayaksterman}.
However, the proof of factorization in heavy quarkonium
production in NRQCD color octet mechanism at all order in coupling constant is still missing.
In this paper we will prove factorization in heavy quarkonium production in NRQCD color octet mechanism at all order
in coupling constant by using path integral method of QCD.

The typical non-perturbative NRQCD matrix element in heavy quarkonium production is given by \cite{nrqcd}
\bea
<0|{\cal O}_n|0> = <0|\chi^\dagger(0) K_n \xi(0) (a^\dagger_H a_H) \xi^\dagger(0) K_n' \chi(0)|0>
\label{bodwin}
\eea
where $\xi$ is the two component Dirac spinor field that annihilates a heavy quark,
$\chi$ is the two component Dirac spinor field that creates a heavy quark,
$a^\dagger_H$ is the operator that creates
the heavy quarkonium $H$ in the out state. The factors $K_n$ and $K_n'$ are products of a color
matrix (either a unit matrix or $T^a$), a spin matrix (either a unit matrix or $\sigma^i$) and
a polynomial of covariant derivative $D$. The color and spin indices on the fields
$\chi$ and $\xi$ have been suppressed.

The production cross section for heavy quarkonium $H$ at transverse momentum $P_T$ in NRQCD factorizes into a
sum of perturbative functions times universal matrix elements,
\bea
d\sigma_{A+B \rightarrow H +X(P_T)} = \sum_n d{\hat \sigma}_{A+B \rightarrow Q{\bar Q}[n] +X(P_T)} ~<{\cal O}_n>
\label{css}
\eea
where each NRQCD non-perturbative matrix element $<{\cal O}_n>$ represents the probability of a heavy quark-antiquark pair in state
[n], such as color singlet or color octet etc., to produce heavy quarkonium state $H$.

The fragmentation function for parton $i$ to evolve into a heavy quarkonium at large $P_T$ is factorized according to
\cite{braatenx}
\bea
D_{H/i}(z, m_c, \mu)= \sum_n d_{i\rightarrow Q{\bar Q}[n]}(z, m_c, \mu) ~<{\cal O}_n>
\label{css1}
\eea
in terms of same NRQCD non-perturbative matrix elements, along with perturbative functions $d_{i\rightarrow Q{\bar Q}[n]}(z, m_c, \mu)$
that describe the evolution of an off-shell parton into a heavy quark-antiquark pair in state
[n], such as color singlet or color octet etc..

At a first glance it can be easily seen that the
non-perturbative NRQCD matrix element in eq. (\ref{bodwin}) is not gauge invariant
unless it is a color singlet S-wave non-perturbative matrix element. Hence one expects that
any non-canceling infrared divergences in the perturbative Feynman diagrams of
heavy quark-antiquark production short-distance coefficient can not be factorized in to the definition of
the non-perturbative NRQCD matrix element in eq. (\ref{bodwin}) to study heavy quarkonium
production at high energy colliders in the NRQCD color octet mechanism.

This is explicitly shown in \cite{nyka,nykb,nayaksterman} where the NNLO coupling constant calculation
shows that the above non-perturbative NRQCD matrix
element in eq. (\ref{bodwin}) is not consistent with factorization of infrared divergences unless it is a color singlet
S-wave non-perturbative matrix element.
By using the calculation at NNLO in coupling constant and to all
powers in the heavy quark relative velocity it was shown in \cite{nyka,nykb,nayaksterman} that
the octet S-wave non-perturbative NRQCD matrix
element which is gauge invariant and is consistent with the factorization of infrared divergences is given by
\bea
<0|{\cal O}_n|0> = <0|\chi^\dagger(0) K_{n,e} \xi(0) \Phi_l^{(A)\dagger}(0)_{eb}(a^\dagger_H a_H) \Phi_l^{(A)}(0)_{ba}\xi^\dagger(0) K'_{n,a} \chi(0)|0>
\label{nrqcdfact}
\eea
where
\bea
\Phi_l^{(A)}(0)={\cal P}{\rm exp}[-igT^{(A)c}\int_0^{\infty} d\lambda l\cdot A^c(l\lambda)],~~~~~~(T^{(A)c})_{ab}=-if^{abc}
\label{adj}
\eea
is the gauge link or the non-abelian phase in the adjoint representation of SU(3), $A^{\mu a}(x)$ is the gluon field,
${\cal P}$ is the path ordering and $l^\mu$ is the light-like four-velocity.

Note that a necessary condition for NRQCD factorization
is that the long-distance behavior of the non-perturbative NRQCD matrix element must be independent of the light-like vector $l^\mu$.
Such a dependence would be inconsistent with NRQCD factorization because the infrared divergences
of $<{\cal O}_n>$ must match those of cross sections, in which there is no information
on $l^\mu$. In \cite{nyka,nykb,nayaksterman} we have verified the $l^\mu$ independence of the infrared pole
at NNLO in coupling constant and to all powers in heavy quark relative velocity.

Since the NRQCD matrix element in eq. (\ref{bodwin}) is a non-perturbative
quantity in QCD it can not be calculated by using perturbative QCD methods.
It is well known that a non-perturbative function can not
be studied by using perturbative methods, no matter how many orders of perturbation theory is used.
Hence path integral formulation (as opposed to diagrammatic method using perturbation theory) is
necessary to study properties of non-perturbative quantities in QCD at all order in coupling constant.
The only path integral formulation to study factorization of soft and collinear divergences at all
order in coupling constant in quantum field theory is given by R. Tucci in \cite{tucci} which is exact for QED but is not
exact for QCD. We have extended this path integral approach to QCD to prove factorization in QCD
at all order in coupling constant in \cite{nayaka3}. In this paper we will extend this path integral approach
to prove NRQCD factorization at all order in coupling constant in heavy quarkonium production.
We will prove that the long-distance behavior of the non-perturbative NRQCD matrix element is
independent of the light-like vector $l^\mu$ at all order in coupling constant.

The paper is organized as follows. In section II we briefly describe the lagrangian density in NRQCD and
in QCD. In section III we discuss infrared divergences in NRQCD and in QCD.
In section IV we include heavy quark in the path integral formulation of QCD.
In section V we describe infrared divergences in NRQCD and the light-like Wilson line in QCD.
In section VI we show that the eikonal current of the light-like charge generates pure gauge
field in quantum field theory.
In section VII we show how the pure gauge field in quantum field theory can be used to describe
soft (infrared) divergences. In section VIII we study heavy quark-antiquark non-perturbative
matrix element in the presence of light-like Wilson line in QCD. In section
IX we prove factorization in heavy quarkonium production in NRQCD color octet mechanism at all order in
coupling constant and to all powers in the heavy quark relative velocity. In section X we show that
the factorization theorem is a key ingredient in calculation of NRQCD heavy quarkonium production cross section
at collider experiments. Section XI contains conclusions.

\section{Lagrangian Density in NRQCD and in QCD }

The lagrangian density in QCD including heavy quarks is given by \cite{muta}
\bea
{\cal L}_{\rm QCD} = -\frac{1}{4}F_{\mu \nu}^aF^{\mu \nu a} + \sum_{l=1}^3 {\bar \psi}_l [\gamma^\mu D_\mu -m_l]\psi_l+{\bar \Psi} [\gamma^\mu D_\mu -M]\Psi
\label{fullqcd}
\eea
where $F^{\mu \nu a}$ is the full non-abelian gluon field tensor, $\psi_l$ is the Dirac field of the
light quark ($l=u,d,s$), $\Psi$ is the Dirac field of the heavy quark, $D^\mu$ is the covariant derivative,
$\gamma^\mu$ is the Dirac matrix, $m_l$ is the mass of the light quark and $M$ is the mass of the heavy quark.

In NRQCD an ultraviolet cutoff $\Lambda \sim M$ is introduced. The lagrangian density in NRQCD is given by \cite{nrqcd}
\bea
{\cal L}_{\rm NRQCD} = {\cal L}_{\rm light}+{\cal L}_{\rm heavy}+\delta{\cal L}
\label{nrqcd}
\eea
where
\bea
{\cal L}_{\rm light} = -\frac{1}{4}F_{\mu \nu}^aF^{\mu \nu a} + \sum_{l=1}^3 {\bar \psi}_l [D -m_l]\psi_l,
\label{nrqcd1}
\eea
\bea
{\cal L}_{\rm heavy} =  \xi^\dagger [iD_t +\frac{{\bf D}^2}{2M}]\xi~+~\chi^\dagger [iD_t -\frac{{\bf D}^2}{2M}]\chi
\label{nrqcd2}
\eea
and
\bea
&&\delta{\cal L}=\frac{c_1}{8M^3} [\xi^\dagger ({\bf D}^2)^2\xi-\chi^\dagger ({\bf D}^2)^2\chi]\nonumber \\
&&+\frac{c_2}{8M^2} [\xi^\dagger ({\bf D} \cdot g{\bf E}-g{\bf E}\cdot {\bf D})\xi-\chi^\dagger ({\bf D} \cdot g{\bf E}-g{\bf E}\cdot {\bf D})\chi]\nonumber \\
&&+\frac{c_3}{8M^2} [\xi^\dagger (i{\bf D} \times g{\bf E}-g{\bf E}\times i{\bf D})\cdot {\bf \sigma} \xi-\chi^\dagger (i{\bf D} \times g{\bf E}-g{\bf E}\times i{\bf D})\cdot {\bf \sigma}\chi]\nonumber \\
&&+\frac{c_4}{2M} [\xi^\dagger g{\bf B}\cdot {\bf \sigma}\xi-\chi^\dagger g{\bf B}\cdot {\bf \sigma}\chi]+...
\label{nrqcd2x}
\eea
where $D_t$ and ${\bf D}$ are the time and space components of the covariant derivative
$D^\mu$ and ${\bf E}$ and ${\bf B}$ are electric and magnetic components of the gluon field tensor and
${\bf \sigma}$ is the Pauli spin matrix. The dimensionless coefficients $c_1$, $c_2$, $c_3$, $c_4$ etc.
in eq. (\ref{nrqcd2x}) are obtained by matching NRQCD with QCD \cite{nrqcd}.

\section{Infrared Behavior in NRQCD and in QCD}

Note that in order for the factorization formula to hold in eqs. (\ref{css}) and (\ref{css1})
the perturbative functions have to be infrared-safe
by definition because infrared limit corresponds to long-distance regime \cite{nrqcd}.
However, as found in \cite{nyka,nykb} the NNLO infrared pole contribution to order ${\vec v}^2$ is given by
\bea
\Sigma(P,v,l) = \alpha_s^2 \frac{1}{3\epsilon} \frac{{\vec v}^2}{4}
\label{pole2}
\eea
which is not zero where ${\vec v}$ is the relative velocity of the heavy quark-antiquark pair. Eq. (\ref{pole2})
is in the rest frame of the heavy quarkonium (${\vec P}=0$)
where $P^\mu$ is the four-momentum of the heavy quarkniuom and $l^\mu$ is the four-velocity of the light-like
Wilson line which is fixed to be $l^\mu =\delta^{\mu -}$ along the minus light cone direction in \cite{nyka,nykb,nayaksterman}.
The presence of non-zero infrared pole in eq. (\ref{pole2}) implies that infrared poles will appear in
perturbative functions at NNLO and beyond when the factorization is carried out with octet
non-perturbative NRQCD matrix element $<\chi^\dagger K_n \xi (a^\dagger_H a_H) \xi^\dagger K'_n \chi>$
in the conventional manner as given by eq. (\ref{bodwin}) in eqs. (\ref{css}) and (\ref{css1}).
On the other hand, when defined according to its gauge-completed form as given by eq. (\ref{nrqcdfact}) each octet
non-perturbative NRQCD matrix element itself generates precisely the same pole terms given in eq. (\ref{pole2})
above. This conclusion is valid to all powers in $v$ at NNLO in coupling constant \cite{nayaksterman}.
Thus NRQCD can accommodate these corrections.
Hence our main aim in this paper is to prove that eq. (\ref{nrqcdfact})
is valid at all order in coupling constant.

Note that in NRQCD an ultraviolet cutoff $\Lambda \sim M$ is introduced \cite{nrqcd}. Hence the
ultraviolet (UV) behavior of QCD and NRQCD differ. However, the
infrared (IR) behavior of QCD and NRQCD remains same \cite{stewart}. Hence the infrared behavior in
NRQCD can be obtained by studying the corresponding infrared behavior in QCD.
Since the matrix element of the type $<\chi^\dagger K_n \xi (a^\dagger_H a_H) \xi^\dagger K'_n \chi>$ is the
non-perturbative NRQCD matrix element, it is natural to study its infrared behavior at all order in coupling
constant by using path integral method.
Hence we will use path integral method of QCD in this paper.

\section{Heavy quarks and the Path integral formulation of QCD }

The generating functional in QCD including the heavy quark is given by \cite{muta,abbott}
\bea
&& Z[J,\eta_u,{\bar \eta}_u,\eta_d,{\bar \eta}_d,\eta_s,{\bar \eta}_s,\eta_h, {\bar \eta}_h]=\int [dQ] [d{\bar \psi}_1] [d \psi_1 ] [d{\bar \psi}_2] [d \psi_2 ][d{\bar \psi}_3] [d \psi_3 ][d{\bar \Psi}] [d \Psi ]~{\rm det}(\frac{\delta (\partial_\mu Q^{\mu a})}{\delta \omega^b}) \nonumber \\
&& e^{i\int d^4x [-\frac{1}{4}{F^a}_{\mu \nu}^2[Q] -\frac{1}{2 \alpha}
(\partial_\mu Q^{\mu a})^2+ J \cdot Q +\sum_{l=1}^3\left[{\bar \psi}_l  [i\gamma^\mu \partial_\mu -m_l +gT^a\gamma^\mu Q^a_\mu] \psi_l +{\bar \eta}_l \psi_l +  {\bar \psi}_l\eta_l\right] +{\bar \Psi}  [i\gamma^\mu \partial_\mu -M +gT^a\gamma^\mu Q^a_\mu] \Psi+{\bar \eta}_h \Psi + {\bar \Psi}\eta_h]}\nonumber \\
\label{aqcd}
\eea
where $Q^{\mu a}$ is the quantum gluon field, the symbols $l=1,2,3=u,d,s$ stand for three light quarks $u,d,s$ and the symbol $h$ stands for
heavy quark and
\bea
F_{\mu \nu}^a[Q]=\partial_\mu Q_\nu^a(x)-\partial_\nu Q_\mu^a(x)+gf^{abc} Q_\mu^b(x)Q_\nu^c(x),~~~~~~~~~~{F^a}_{\mu \nu}^2[Q]=F_{\mu \nu}^a[Q]F^{\mu \nu a}[Q].\nonumber \\
\eea
In eq. (\ref{aqcd}) the ${\bar \eta}_u,~{\bar \eta}_d,~{\bar \eta}_s$ are external sources for $u,d,s$ quark fields
respectively and ${\bar \eta}_h$ is the external source for the heavy quark field and the term
$\frac{\delta (\partial_\mu Q^{\mu a})}{\delta \omega^b}$
is the derivative of the gauge fixing term under an infinitesimal gauge transformation \cite{muta,abbott}
\bea
\delta Q^{\mu a} =gf^{abc}Q^{\mu b}\omega^c+\partial^\mu \omega^a.
\label{aqcdgf}
\eea
Note that the determinant ${\rm det}(\frac{\delta (\partial_\mu Q^{\mu a})}{\delta \omega^b})$ in eq. (\ref{aqcd}) can be
expressed in terms of path integration over the ghost fields \cite{muta}. However, we will directly work with the
determinant ${\rm det}(\frac{\delta (\partial_\mu Q^{\mu a})}{\delta \omega^b})$ in eq. (\ref{aqcd}).

For the heavy quark Dirac field $\Psi(x)$,
the non-perturbative matrix element of the type $<0|{\bar \Psi}(x) O_n \Psi(x) {\bar \Psi}(x') O'_n \Psi(x')|0>$ in
QCD is given by \cite{tucci}
\bea
&&<0|{\bar \Psi}(x) O_n \Psi(x) {\bar \Psi}(x') O'_n \Psi(x')|0>\nonumber \\
&&=\frac{\delta}{\delta \eta_h(x)}O_n\frac{\delta}{\delta {\bar \eta}_h(x)}\frac{\delta}{\delta \eta_h(x')}O'_n\frac{\delta}{\delta {\bar \eta}_h(x')}Z[J,\eta_u,{\bar \eta}_u,\eta_d,{\bar \eta}_d,\eta_s,{\bar \eta}_s,\eta_h, {\bar \eta}_h]|_{J=\eta_u={\bar \eta}_u=\eta_d={\bar \eta}_d=\eta_s={\bar \eta}_s=\eta_h= \eta_h=0}\nonumber \\
\label{alepg}
\eea
if the factors $O_n$ and $O'_n$ are independent of quantum fields where the suppression of the normalization factor $Z[0]$
is understood as it will cancel in the final result (see eq. (\ref{finalfacts})).

\section{Infrared Divergences in NRQCD and Light-Like Wilson Line in QCD}

The gauge transformation of the quark field in QCD is given by
\bea
\psi'(x)=e^{igT^a\omega^a(x)}\psi(x).
\label{phg3}
\eea
Hence one finds that the issue of gauge invariance and factorization of infrared divergences in QCD can be simultaneously
explained if $\omega^a(x)$ can be related to the gluon field $ {A}^{\mu a}(x)$.

Before proceeding to the issue of gauge invariance and the factorization of infrared divergences in QCD let us first discuss
the corresponding situation in QED. The gauge transformation of the Dirac field of the electron in QED is given by
\bea
\psi'(x)=e^{ie\omega(x)}\psi(x).
\label{phg3q}
\eea
Hence we can expect to address the issue of gauge invariance and factorization of infrared divergences
in QED simultaneously if we can relate the $\omega(x)$ to the photon field $A^\mu(x)$.

In QED the infrared (or soft) divergence arises only from the emission of a photon for which all components of the four-momentum
are small. The Eikonal propagator times the Eikonal vertex for a soft photon with momentum $k$ interacting with a
light-like electron moving with four momentum $p^\mu$ is given by
\cite{collins,tucci,collinssterman,berger,bodwin,frederix,nayakqed,scet1,pathorder,nayaka2,nayaka3}
\bea
e~\frac{p^\mu}{p \cdot k+i\epsilon }=e~\frac{l^\mu}{l \cdot k+ i\epsilon }
\label{eikonaliq}
\eea
where $l^\mu$ is the four-velocity of the light-like electron. Note that when we say the "light-like electron" we mean the electron
that is traveling at its highest speed which is arbitrarily close to the speed of light
($|{\vec l}|\sim 1$) as it can not travel exactly at speed of light ($|{\vec l}|= 1$)
because it has finite mass even if the mass of the electron is very small. From eq. (\ref{eikonaliq})
we find
\bea
&& e\int \frac{d^4k}{(2\pi)^4} \frac{l\cdot {  A}(k)}{l\cdot k +i\epsilon } =-e i\int_0^{\infty} d\lambda \int \frac{d^4k}{(2\pi)^4} e^{i l \cdot k \lambda} l\cdot {A}(k) = ie\int_0^{\infty} d\lambda l\cdot { A}(l\lambda)
\label{ftgtq}
\eea
where the photon field  $ { A}^{\mu }(x)$ and its Fourier transform $ { A}^{\mu }(k)$ are related by
\bea
{ A}^{\mu }(x) =\int \frac{d^4k}{(2\pi)^4} { A}^{\mu }(k) e^{ik \cdot x}.
\label{ftq}
\eea
From eq. (\ref{ftgtq}) we find
\bea
ie\int_0^{\infty} d\lambda l\cdot { A}(l\lambda)=i \int d^4x J^\mu(x) A_\mu(x)
\label{ftgtqmn}
\eea
where the eikonal current density $J^\mu(x)$ for the light-like charge $e$ is given by
\bea
J^\mu(x) = el^\mu \int d\lambda \delta^{(4)}(x-l\lambda).
\label{ekcd}
\eea

Now consider the corresponding Feynman diagram for the infrared divergences in QED
due to exchange of two soft-photons of four-momenta $k^\mu_1$ and $k^\mu_2$.
The corresponding Eikonal contribution due to two soft-photons exchange is analogously given by
\bea
&& e^2\int \frac{d^4k_1}{(2\pi)^4} \frac{d^4k_2}{(2\pi)^4} \frac{ l\cdot { A}(k_2) l\cdot { A}(k_1)}{(l\cdot (k_1+k_2) +i \epsilon)(l\cdot k_1 +i \epsilon)} \nonumber \\
&&=e^2i^2 \int_0^{\infty} d\lambda_2 \int_{\lambda_2}^{\infty} d\lambda_1 l\cdot { A}(l\lambda_2) l\cdot {\cal A}(l\lambda_1)
\nonumber \\
&&= \frac{e^2i^2}{2!}\int_0^{\infty} d\lambda_2 \int_0^{\infty} d\lambda_1 l\cdot {\cal A}(l\lambda_2) l\cdot {\cal A}(l\lambda_1).
\eea
Extending this calculation up to infinite number of soft-photons
we find that the Eikonal contribution for the infrared divergences due to
soft photons exchange with the light-like electron in QED is given by the exponential
\bea
e^{ie \int_0^{\infty} d\lambda l\cdot { A}(l\lambda) }
\label{iiijqed}
\eea
where $l^\mu$ is the light-like four velocity of the electron. The Wilson line in QED is given by
\bea
e^{ie \int_{x_i}^{x_f} dx^\mu A_\mu(x) }.
\label{klj}
\eea
When $A^\mu(x)=A^\mu(l\lambda)$ as in eq. (\ref{iiijqed}) then
one finds from eq. (\ref{klj}) that the light-like Wilson line in QED for infrared divergences is given by \cite{stermanpath}
\bea
e^{ie \int_0^x dx^\mu A_\mu(x) }=e^{-ie \int_0^{\infty} d\lambda l\cdot { A}(x+l\lambda) }e^{ie \int_0^{\infty} d\lambda l\cdot { A}(l\lambda) }.
\label{tto}
\eea
Note that a light-like electron traveling with light-like four-velocity $l^\mu$ produces U(1) pure gauge potential $A^{\mu }(x)$
at all the time-space position $x^\mu$ except at the position ${\vec x}$ perpendicular to the direction of motion
of the electron (${\vec l}\cdot {\vec x}=0$) at the time of closest approach \cite{collinssterman,nayakj,nayake}.
When $A^{\mu }(x) = A^{\mu }(\lambda l)$ as in eq. (\ref{iiijqed})
we find ${\vec l}\cdot {\vec x}=\lambda {\vec l}\cdot {\vec l}=\lambda\neq 0$ which implies that the light-like Wilson line
finds the photon field $A^{\mu }(x)$ in eq. (\ref{iiijqed}) as the U(1) pure gauge. The U(1) pure gauge is given by
\bea
A^\mu(x)=\partial^\mu \omega(x)
\label{purea1}
\eea
which gives from eq. (\ref{tto}) the light-like Wilson line in QED for infrared divergences
\bea
e^{ie\omega(x)}e^{-ie\omega(0)}=e^{ie \int_0^x dx^\mu A_\mu(x) }=e^{-ie \int_0^{\infty} d\lambda l\cdot { A}(x+l\lambda) }e^{ie \int_0^{\infty} d\lambda l\cdot { A}(l\lambda) }
\label{lkj}
\eea
which depends only on end points $0$ and $x^\mu$ but is independent of the path.
The path independence can also be found from Stokes theorem because for pure gauge
\bea
F^{\mu \nu}(x)=\partial^\mu A^\nu(x)-\partial^\nu A^\mu(x)=0
\eea
which gives from Stokes theorem
\bea
e^{ie \oint_C dx^\mu A_\mu(x) }=e^{ie \int_S dy^\mu dx^\nu F_{\mu \nu}(x) }=1
\eea
where $C$ is a closed path and $S$ is the surface enclosing $C$. Now considering two different paths
$L$ and $M$ with common end points $0$ and $x^\mu$ we find
\bea
e^{ie \oint_C dx^\mu A_\mu(x) }=e^{ie \int_L dx^\mu A_\mu(x)- ie \int_M dx^\mu A_\mu(x)}=1
\eea
which implies that
\bea
e^{ie \int_0^x dx^\mu A_\mu(x) }
\eea
depends only on end points $0$ and $x^\mu$ but is independent of path which can also be seen from eq. (\ref{lkj}).
Hence from eq. (\ref{lkj}) we find that the abelian phase or the gauge link in QED is given by
\bea
e^{-ie \int_0^{\infty} d\lambda l\cdot { A}(x+l\lambda) }=e^{ie\omega(x)}.
\label{phas}
\eea
From eqs. (\ref{phg3q}) and (\ref{phas}) one expects that the gauge invariance and factorization of infrared divergences
in QED can be explained simultaneously.

One can recall that the gauge invariant greens function in QED
\bea
G(x_1,x_2)=<0|{\bar \psi}(x_2)~\times~{\rm exp}[ie \int_{x_1}^{x_2} dx^\mu A_\mu(x)]~\times~ \psi(x_1)|0>
\label{qed}
\eea
in the presence of background field $A^\mu(x)$ was formulated by Schwinger long time ago \cite{schw1}.
When this background field $A^\mu(x)$ is replaced by the U(1) pure gauge background field
as given by eq. (\ref{purea1}) then one finds by using the path integral method of QED that \cite{tucci}
\bea
&&e^{ie\omega(x_2)}<0|{\bar \psi}(x_2)~ \psi(x_1)|0>_A e^{-ie\omega(x_1)}= <0|{\bar \psi}(x_2)~ \psi(x_1)|0>\nonumber \\
&&=e^{-ie\int_0^{\infty} d\lambda l\cdot { A}(x_2+l\lambda)}<0|{\bar \psi}(x_2)~ \psi(x_1)|0>_A e^{ie\int_0^{\infty} d\lambda l\cdot { A}(x_1+l\lambda)}
\label{tucci}
\eea
which proves the gauge invariance and factorization
of infrared divergences in QED simultaneously.
In eq. (\ref{tucci}) the $<0|{\bar \psi}(x_2)~ \psi(x_1)|0>$ is the full Green's function
in QED and $<0|{\bar \psi}(x_2)~ \psi(x_1)|0>_A$ is the corresponding Green's function in
the background field method of QED. This path integral technique is also used in \cite{nayakqed} to prove factorization of
infrared divergences in non-equilibrium QED.

Hence we find that the gauge invariance and factorization of infrared divergences in QED can be studied by using
path integral method of QED in the presence of U(1) pure gauge background field.
Therefore one expects that the gauge invariance and factorization of infrared divergences in QCD can be studied by using
path integral method of QCD in the presence of SU(3) pure gauge background field.

Now let us proceed to QCD.
In QCD the infrared (or soft) divergence arises only from the emission of a gluon for which all components of the four-momentum
are small. The Eikonal propagator times the Eikonal vertex for a soft gluon with momentum $k$ interacting with a light-like
quark moving with four momentum $p^\mu$ is given by
\cite{collins,tucci,collinssterman,berger,frederix,nayakqed,scet1,pathorder,nayaka2,nayaka3}
\bea
gT^a~\frac{p^\mu}{p \cdot k+i\epsilon }=gT^a~\frac{l^\mu}{l \cdot k+ i\epsilon }
\label{eikonalin}
\eea
where $l^\mu$ is the four-velocity of the light-like quark. Note that when we say the "light-like quark" we mean the quark
that is traveling at its highest speed which is arbitrarily close to the speed of light ($|{\vec l}|\sim 1$)
as it can not travel exactly at speed of light ($|{\vec l}|= 1$)
because it has finite mass even if the mass of the light quark is very small. On the other hand the gluon is massless and hence it always
travels at speed of light and is exactly light-like. From eq. (\ref{eikonalin}) we find
\bea
&& gT^a\int \frac{d^4k}{(2\pi)^4} \frac{l\cdot { A}^a(k)}{l\cdot k +i\epsilon } =-gT^a i\int_0^{\infty} d\lambda \int \frac{d^4k}{(2\pi)^4} e^{i l \cdot k \lambda} l\cdot { A}^a(k) = igT^a\int_0^{\infty} d\lambda l\cdot {  A}^a(l\lambda)\nonumber \\
\label{ftgt}
\eea
where the gluon field $ { A}^{\mu a}(x)$ and its Fourier transform $ { A}^{\mu a}(k)$ are related by
\bea
{ A}^{\mu a}(x) =\int \frac{d^4k}{(2\pi)^4} { A}^{\mu a}(k) e^{ik \cdot x}.
\label{ft}
\eea
Note that a path ordering in QCD is required which can be seen as follows, see also \cite{bodwin}. The
Eikonal contribution for the infrared divergence in QCD arising from a single soft-gluon exchange in Feynman diagram
is given by eq. (\ref{ftgt}). Now consider the corresponding Feynman diagram for the infrared divergences in QCD
due to exchange of two soft-gluons of four-momenta $k^\mu_1$ and $k^\mu_2$.
The corresponding Eikonal contribution due to two soft-gluons exchange is analogously given by
\bea
&& g^2\int \frac{d^4k_1}{(2\pi)^4} \frac{d^4k_2}{(2\pi)^4} \frac{T^a l\cdot { A}^a(k_2)T^b l\cdot { A}^b(k_1)}{(l\cdot (k_1+k_2) +i \epsilon)(l\cdot k_1 +i \epsilon)} \nonumber \\
&&=g^2i^2 \int_0^{\infty}  d\lambda_2 \int_{\lambda_2}^{\infty} d\lambda_1 T^a l\cdot { A}^a(l\lambda_2) T^b l\cdot { A}^b(l\lambda_1)
\nonumber \\
&&= \frac{g^2i^2}{2!} {\cal P}\int_0^{\infty}  d\lambda_2 \int_0^{\infty}  d\lambda_1 T^a l\cdot { A}^a(l\lambda_2) T^b l\cdot { A}^b(l\lambda_1)
\eea
where ${\cal P}$ is  the path ordering.
Extending this calculation up to infinite number of soft-gluons we find that the Eikonal contribution for the infrared
divergences due to soft gluons exchange with the light-like quark in QCD is given by the path ordered exponential
\bea
{\cal P}~{\rm exp}[ig \int_0^{\infty} d\lambda l\cdot { A}^a(l\lambda)T^a ]
\label{iiij}
\eea
where $l^\mu$ is the light-like four velocity of the quark. The Wilson line in QCD is given by
\bea
{\cal P}e^{ig \int_{x_i}^{x_f} dx^\mu A_\mu^a(x)T^a }
\label{tts}
\eea
which is the solution of the equation \cite{lam}
\bea
\partial_\mu S(x)=igT^aA_\mu^a(x)S(x)
\eea
with initial condition
\bea
S(x_i)=1.
\eea
When $A^{\mu a}(x)=A^{\mu a}(l\lambda)$ as in eq. (\ref{iiij}) we find from eq. (\ref{tts}) that the light-like Wilson line in
QCD for infrared divergences is given by \cite{stermanpath}
\bea
{\cal P}e^{ig \int_0^x dx^\mu A_\mu^a(x) T^a}=\left[{\cal P}e^{-ig \int_0^{\infty} d\lambda l\cdot { A}^a(x+l\lambda) T^a}\right]{\cal P}e^{ig \int_0^{\infty} d\lambda l\cdot { A}^b(l\lambda) T^b}.
\label{oh}
\eea

A light-like quark traveling with light-like four-velocity $l^\mu$ produces SU(3) pure gauge potential $A^{\mu a}(x)$
at all the time-space position $x^\mu$ except at the position ${\vec x}$ perpendicular to the direction of motion
of the quark (${\vec l}\cdot {\vec x}=0$) at the time of closest approach \cite{collinssterman,nayakj,nayake}.
When $A^{\mu a}(x) = A^{\mu a}(\lambda l)$ as in eq. (\ref{iiij})
we find ${\vec l}\cdot {\vec x}=\lambda {\vec l}\cdot {\vec l}=\lambda\neq 0$ which implies that the light-like Wilson line
finds the gluon field $A^{\mu a}(x)$ in eq. (\ref{iiij}) as the SU(3) pure gauge. The SU(3) pure gauge is given by
\bea
T^aA_\mu^a (x)= \frac{1}{ig}[\partial_\mu U(x)] ~U^{-1}(x),~~~~~~~~~~~~~U(x)=e^{igT^a\omega^a(x)}
\label{gtqcd}
\eea
which gives
\bea
U(x_f)={\cal P}e^{ig \int_{x_i}^{x_f} dx^\mu A_\mu^a(x) T^a}U(x_i)=e^{igT^a\omega^a(x_f)}.
\label{uxf}
\eea
Hence when $A^{\mu a}(x) = A^{\mu a}(\lambda l)$ as in eq. (\ref{iiij}) we find from eqs. (\ref{oh}) and
(\ref{uxf}) that the light-like Wilson line in QCD for infrared divergences is given by
\bea
{\cal P}e^{ig \int_{0}^{x} dx^\mu A_\mu^a(x)T^a }=e^{igT^a\omega^a(x)}e^{-igT^b\omega^b(0)}=\left[{\cal P}e^{-ig \int_{0}^{\infty} d\lambda l\cdot { A}^a(x+l\lambda)T^a }\right]{\cal P}e^{ig \int_{0}^{\infty} d\lambda l\cdot { A}^b(l\lambda)T^b }
\label{lkjn}
\eea
which depends only on end points $0$ and $x^\mu$ but is independent of the path.
The path independence can also be found from the non-abelian Stokes theorem which can be seen as follows.
The SU(3) pure gauge in eq. (\ref{gtqcd})
gives
\bea
F^a_{\mu \nu}[A]=\partial_\mu A^a_\nu(x) - \partial_\nu A^a_\mu(x)+gf^{abc} A^b_\mu(x) A^c_\nu(x)=0.
\label{cfmn}
\eea
Note that from eq. (\ref{cfmn}) we find the vanishing physical gauge invariant field strength square $F^{\mu \nu a}[A]F^a_{\mu \nu}[A]$
when $A^{\mu a}(x)$ is the SU(3) pure gauge as given by eq. (\ref{gtqcd}).
Hence in classical mechanics the SU(3) pure gauge potential does not have an effect on color charged
particle and one expects the effect of exchange of soft gluons to simply vanish.
However, in quantum mechanics the situation is a little more complicated, because the gauge potential
does have an effect on color charged particle even if it is SU(3) pure gauge potential and hence
one should not expect the effect of exchange of soft gluons to simply vanish \cite{collinssterman}.
This can be verified by studying the non-perturbative matrix element
in QCD such as $<{\bar \Psi}(x) \Psi(x') {\bar \Psi}(x'') \Psi(x''')...>$ in the
presence of SU(3) pure gauge background field.

Using eq. (\ref{cfmn}) in the non-abelian Stokes theorem \cite{stokes} we find
\bea
{\cal P}e^{ig \oint_C dx^\mu A_\mu^a(x)T^a} = {\cal P}{\rm exp}[ig \int_S dx^\mu dx^\nu \left[{\cal P}e^{ig \int_y^x dx'^\lambda A_\lambda^b(x')T^b}\right]F_{\mu \nu}^a(x)  T^a\left[{\cal P}e^{ig \int_x^y dx''^\delta A_\delta^c(x'')T^c}\right]]=1\nonumber \\
\label{555}
\eea
where $C$ is a closed path and $S$ is the surface enclosing $C$. Now considering two different paths
$L$ and $M$ with common end points $0$ and $x^\mu$ we find from eq. (\ref{555})
\bea
&&{\cal P}e^{ig \oint_C dx^\mu A_\mu^a(x)T^a} ={\cal P}{\rm exp}[ig \int_L dx^\mu A_\mu^a(x)T^a- ig \int_M dx^\mu A_\mu^a(x)T^a]\nonumber \\
&&=\left[{\cal P}e^{ig \int_L dx^\mu A_\mu^a(x)T^a}\right]\left[{\cal P}e^{- ig \int_M dx^\nu A_\nu^b(x)T^b}\right]=1
\label{pht}
\eea
which implies that the light-like Wilson line in QCD
\bea
{\cal P}e^{ig \int_0^x dx^\mu A_\mu^a(x)T^a }
\label{llwl}
\eea
depends only on the end points $0$ and $x^\mu$ but is independent of the path which can also be seen from eq. (\ref{lkjn}).
Hence from eq. (\ref{lkjn}) we find that the non-abelian phase or the gauge link in QCD is given by
\bea
\Phi(x)={\cal P}e^{-ig \int_0^{\infty} d\lambda l\cdot { A}^a(x+l\lambda)T^a }=e^{igT^a\omega^a(x)}.
\label{ttt}
\eea
In the adjoint representation of SU(3) the corresponding path ordered exponential is given by
\bea
{\cal P}{\rm exp}[-ig\int_0^{\infty} d\lambda l\cdot { A}^c(x+l\lambda)T^{(A)c}]=e^{igT^{(A)c}\omega^c(x)},~~~~~~(T^{(A)c})_{ab}=-if^{abc}.
\label{adjin}
\eea

To summarize this, we find that the infrared divergences in the perturbative Feynman diagrams due to soft-gluons interaction
with the light-like Wilson line in QCD is given by the path ordered exponential in eq. (\ref{iiij}) which is nothing
but the non-abelian phase or the gauge link in QCD as given by eq. (\ref{ttt}) where the gluon field $A^{\mu a}(x)$ is the
SU(3) pure gauge, see eqs. (\ref{gtqcd}), (\ref{uxf}), (\ref{lkjn}). This implies that the effect of soft-gluons
interaction between the partons and the light-like Wilson line in QCD can be studied by
putting the partons in the SU(3) pure gauge background field.
Hence we find that the infrared behavior of the non-perturbative matrix element
such as $<0|{\bar \Psi}(x) \Psi(x') {\bar \Psi}(x'') \Psi(x''')...|0>$ in QCD
due to the presence of light-like Wilson line in QCD can be studied by using the path integral method
of the QCD in the presence of SU(3) pure gauge background field.

It can be mentioned here that in soft collinear effective theory
(SCET) \cite{scet} it is also necessary to use the idea of background fields \cite{abbott} to give well defined meaning to several
distinct gluon fields \cite{scet1}.

As mentioned earlier,
in NRQCD an ultraviolet cutoff $\Lambda \sim M$ is introduced \cite{nrqcd}. Hence the
ultraviolet (UV) behavior of QCD and NRQCD differ. However, the
infrared (IR) behavior of QCD and NRQCD remains same \cite{stewart}. Hence the infrared behavior in
NRQCD can be studied by studying the corresponding infrared behavior in QCD. Hence we find that
the infrared behavior of the non-perturbative NRQCD matrix element
$<0|\chi^\dagger K_n \xi (a^\dagger_H a_H) \xi^\dagger K_n' \chi|0>$ in eq. (\ref{bodwin})
can be obtained by studying the infrared behavior of the non-perturbative matrix element in QCD of the type
$<0|{\bar \Psi}(x) O_n \Psi(x) {\bar \Psi}(x') O'_n \Psi(x')|0>$
where $O_n,~O'_n$ are appropriate factors which identify the state of the
heavy quark-antiquark system such as the color singlet state or color octet state etc..

Note that a massive color source traveling at speed much less than speed of light
can not produce SU(3) pure gauge field \cite{collinssterman,nayakj,nayake}. Hence when one replaces light-like
Wilson line with massive Wilson line one expects the factorization of infrared divergences to
break down. This is in confirmation with the finding in \cite{nayaksterman1} which used the diagrammatic
method of QCD. In case of massive Wilson line in QCD the color transfer occurs and the factorization breaks
down.

\section{ Eikonal Current of the Light-Like Charge Generates Pure Gauge Field in Quantum Field Theory }

In order to study factorization of infrared divergences by using the background field method of QED,
the soft photon cloud traversed by the electron is represented by the pure gauge background field
$A^\mu(x)$ \cite{tucci} due to the presence of the light-like Wilson line, where one represents the
quantum photon field by $Q^\mu(x)$. As mentioned above, in classical mechanics the assertion that the gauge field that is produced
by a highly relativistic (light-like) particle is a pure gauge \cite{collinssterman,nayakj,nayake}.
One may ask a question if this assertion is correct in quantum field theory. In this section
we will show that this assertion is correct in quantum field theory. We will use path integral formulation
of the quantum field theory for this purpose.

The generating functional for the gauge field
in the quantum field theory in the presence of external source $J^\mu(x)$ in the path integral formulation is given by
\bea
Z[J]=\int [dQ]
e^{i\int d^4x [-\frac{1}{4}{F}_{\mu \nu}^2[Q] -\frac{1}{2 \alpha} (\partial_\mu Q^{\mu })^2+ J \cdot Q ]}
\label{zfqv}
\eea
where $Q^\mu(x)$ is the quantum photon field and
\bea
F^{\mu \nu}[Q]=\partial^\mu Q^\nu(x)-\partial^\nu Q^\mu(x),~~~~~~~~~{F}_{\mu \nu}^2[Q]={F}^{\mu \nu}[Q]{F}_{\mu \nu}[Q].
\eea
The effective action $S_{eff}[J]$ is given by \cite{peter}
\bea
<0|0>_J =\frac{Z[J]}{Z[0]}=e^{iS_{eff}[J]}
\label{vacv}
\eea
where
\bea
S_{eff}[J] = -\frac{1}{2} \int d^4x d^4x' J^\mu(x) D_{\mu \nu}(x-x')J^\nu(x')
\label{wja}
\eea
where $D_{\mu \nu}(x-x')$ is the photon propagator.

The photon propagator in the coordinate space is given by
\bea
D_{\mu \nu}(x-x') = \frac{1}{\partial^2}[g_{\mu \nu}+\frac{(\alpha -1)}{\partial^2}\partial_\mu \partial_\nu] \delta^{(4)}(x-x').
\label{dmnc}
\eea
Using eq. (\ref{dmnc}) in (\ref{wja}) we find
\bea
S_{eff}[J] = -\frac{1}{2} \int d^4x J^\mu(x)\frac{1}{\partial^2} [g_{\mu \nu}+\frac{(\alpha -1)}{\partial^2}\partial_\mu \partial_\nu] J^\nu(x).
\label{wjh}
\eea
From the continuity equation we have
\bea
\partial_\mu J^\mu(x)=0.
\label{ceq}
\eea
Using eq. (\ref{ceq}) in (\ref{wjh}) we find
\bea
S_{eff}[J] = -\frac{1}{2} \int d^4x J^\mu(x)  \frac{1}{\partial^2}J_\mu(x).
\label{wj}
\eea

\subsection{Derivation of Coulomb's Law For Static Charge in Quantum Field Theory }

First of all, by using the path integral formulation of the quantum field theory we will derive Coulomb's
law for static charge. Note that the derivation of the Coulomb's law by using path integral
formulation of the quantum field theory is not necessary to prove factorization theorem.
We have included it here only to demonstrate the correctness of the prediction of the
path integral formulation in quantum field theory which we will use (see below)
to show that the eikonal current of the light-like charge generates pure gauge field
in quantum field theory.

In order to derive Coulomb's law by using path integral formulation
of the quantum field theory we consider two static charges
at positions ${\vec X}$ and ${\vec X}'$ respectively. The current density for this two static
charges is given by
\bea
J^\mu(x) =e\delta^{\mu 0} \delta^{(3)}({\vec x}-{\vec X})+e\delta^{\mu 0} \delta^{(3)}({\vec x}-{\vec X}').
\label{cdr}
\eea
Using eq. (\ref{cdr}) in (\ref{wj}) and neglecting the self energies we find in the time interval $t$ that
\bea
S_{eff}^{int}[J] = e^2 \int dt \frac{1}{\nabla^2_X}\delta^{(3)}({\vec X}-{\vec X}')=-te^2\frac{1}{\nabla^2_X}\nabla_X^2(\frac{1}{|{\vec X}-{\vec X}'|})= -t V_{eff}^{int}[J]
\label{wjr}
\eea
which gives the (effective) potential energy $V_{eff}[J]$ of the interaction between two static charges to be
\bea
V_{eff}^{int}[J]=\frac{e^2}{|{\vec X}-{\vec X}'|}
\eea
which reproduces the Coulomb's law. Hence we have shown that the assertion that a charge
at rest generates a Coulomb gauge field is correct in quantum field theory.

\subsection{ Effective Lagrangian Density of Light-Like Eikonal Current in Quantum Field Theory }

Similarly using the above procedure in quantum field theory we will show that the assertion that a light-like
charge generates pure gauge field is correct in quantum field theory. This can be shown
as follows.

The eikonal current density of the charge $e$ with light-like four-velocity $l^\mu$ is given by eq. (\ref{ekcd}).
By using the path integral formulation of the quantum field theory we find by using eq.
(\ref{ekcd}) in (\ref{wj}) that for light-like eikonal current the effective lagrangian density is given by
\bea
{\cal L}_{eff}(x)=\frac{e^2}{2} \frac{(l^2)^2 }{(l \cdot x)^4}.
\label{efefi}
\eea
For light-like four-velocity we have
\bea
l^2=l^\mu l_\mu=0.
\label{lsq2}
\eea
Hence from (\ref{efefi}) and (\ref{lsq2}) we find that for light-like eikonal current the effective lagrangian density is given by
\bea
{\cal L}_{eff}(x)=0, ~~~~~~~~~~~~~~~l\cdot x \neq 0
\label{eeffl}
\eea
at all the time-space position $x^\mu$ except at the spatial position perpendicular to the motion of the charge $({\vec l} \cdot {\vec x}=0)$
at the time of closest approach $(x_0=0)$.

\subsection{ Interaction Between Non-Eikonal Current and the Gauge Field Generated by Light-Like Eikonal Current in Quantum Field Theory}

Similarly by using the above path integral formulation calculation we find from eq. (\ref{wjf}) that the interaction
between the (light-like or non-light-like) non-eikonal current and the gauge field generated by
the light-like eikonal current gives the effective (interaction) lagrangian density
\bea
{\cal L}^{int}_{eff}(x)=l^2 \frac{e^2}{2}\frac{(l \cdot q) (q \cdot x) -(l \cdot x) q^2 }{(l \cdot x)^3[(q \cdot x)^2 -q^2 x^2]^{\frac{3}{2}}}
\label{nonef}
\eea
where $q^\mu$ is the (light-like or non-light-like) four-momentum of non-eikonal current of charge $e$ and
$l^\mu$ is the light-like four-velocity of the eikonal current of charge $e$.

For light-like eikonal current we find from eqs. (\ref{lsq2}) and (\ref{nonef}) that effective (interaction) lagrangian density
due to the interaction between the (light-like or non-light-like) non-eikonal current of four-momentum $q^\mu$ and the gauge
field generated by the light-like eikonal current of four-velocity $l^\mu$ is given by
\bea
{\cal L}^{int}_{eff}(x)=0,~~~~~~~~~~~~{\rm for}~~~~~~~~~~~q\cdot x \neq 0,~~~~~~~~~~l\cdot x  \neq 0.
\label{ffl}
\eea
This is also obvious from eq. (\ref{totalf}).

\subsection{Pure Gauge Field Generated By Eikonal Current of Light-Like Charge in Quantum Field Theory }

Hence from eqs. (\ref{eeffl}) and (\ref{ffl}) we find that the eikonal current for
light-like charge generates pure gauge field in quantum field theory.
From eqs. (\ref{eeffl}) and (\ref{ffl}) we find that the assertion
that a light-like charge generates a pure gauge field is correct in quantum field theory
which is consistent with the corresponding result in classical mechanics \cite{collinssterman,nayakj,nayake}.

\section{ Pure Gauge Field in Quantum Field Theory Describes Soft (Infrared) Divergence }

In this section we will show how the pure gauge field is used in quantum field theory to describe
soft (infrared) divergences.
Consider an incoming electron of four momentum $q^\mu$ and mass $m$ emitting a real photon of four momentum $k^\mu$.
The corresponding Feynman diagram contribution is given by \cite{grammer}
\begin{eqnarray}
&& {\cal M}=\frac{1}{\gamma_\nu q^\nu -\gamma_\nu k^\nu -m} \gamma_\mu \epsilon^\mu(k)u(q)=-\frac{q \cdot \epsilon(k)}{q \cdot k}u(q)+\frac{k^\nu \gamma_\nu \gamma_\mu \epsilon^\mu(k)}{2q \cdot k}u(q)
\label{total}
\end{eqnarray}
where we write
\begin{eqnarray}
{\cal M}_{\rm eikonal}=-\frac{q \cdot \epsilon(k)}{q \cdot k}u(q)
\label{eik}
\end{eqnarray}
and
\begin{eqnarray}
{\cal M}_{\rm non-eikonal}=\frac{k^\nu \gamma_\nu \gamma_\mu \epsilon^\mu(k)}{2q \cdot k} u(q).
\label{noneik}
\end{eqnarray}
From eq. (3.2) of \cite{grammer} we write the gauge field as
\begin{equation}
\epsilon^\mu(k) = [\epsilon^\mu(k) -k^\mu \frac{q \cdot \epsilon(k)}{q \cdot k}]+k^\mu \frac{q \cdot \epsilon(k)}{q \cdot k}=\epsilon_{\rm phys}^\mu(k)+\epsilon_{\rm pure}^\mu(k)
\label{1}
\end{equation}
where
\begin{equation}
\epsilon_{\rm phys}^\mu(k) = [\epsilon^\mu(k) -k^\mu \frac{q \cdot \epsilon(k)}{q \cdot k}]
\label{1a}
\end{equation}
is the physical gauge field [corresponding to transverse polarization of the gauge field] and
\begin{equation}
\epsilon_{\rm pure}^\mu(k) = k^\mu \frac{q \cdot \epsilon(k)}{q \cdot k}
\label{1b}
\end{equation}
is the pure gauge field [corresponding to longitudinal polarization of the gauge field].

Now using eq. (\ref{1}) in eq. (\ref{total}) we find that the total contribution of the Feynman diagram is given by
\begin{eqnarray}
&& {\cal M}= {\cal M}_{\rm eikonal}+{\cal M}_{\rm non-eikonal}
\label{totala}
\end{eqnarray}
where
\begin{eqnarray}
{\cal M}_{\rm eikonal} = -\frac{q \cdot \epsilon_{\rm phys}(k)}{q \cdot k}u(q) - \frac{q \cdot \epsilon_{\rm pure}(k)}{q \cdot k}u(q)=-\frac{q \cdot \epsilon_{\rm pure}(k)}{q \cdot k}u(q)
\label{totalb}
\end{eqnarray}
and
\begin{eqnarray}
{\cal M}_{\rm non-eikonal} = \frac{k^\nu \gamma_\nu \gamma_\mu \epsilon_{\rm phys}^\mu(k)}{2q \cdot k} u(q)+\frac{k^\nu \gamma_\nu \gamma_\mu \epsilon_{\rm pure}^\mu(k)}{2q \cdot k} u(q)=\frac{k^\nu \gamma_\nu \gamma_\mu \epsilon_{\rm phys}^\mu(k)}{2q \cdot k} u(q).
\label{totalc}
\end{eqnarray}

Hence in the soft photon limit $(k_0,k_1,k_2,k_3) \rightarrow 0$ we find from the eqs. (\ref{total}) and (\ref{totalb}) that
\begin{eqnarray}
-{\cal M}_{\rm eikonal}=\frac{q \cdot \epsilon(k)}{q \cdot k}u(q) =\frac{q \cdot \epsilon_{\rm pure}(k)}{q \cdot k}u(q) \rightarrow \infty ~~~~~~~~~~~{\rm as}~~~~~~~~~~(k_0,k_1,k_2,k_3) \rightarrow 0
\label{totald}
\end{eqnarray}
which implies that the physical gauge field [corresponding to transverse polarization] does not contribute to
the soft (infrared) divergences in quantum field theory and the soft (infrared) divergences can be calculated
by using pure gauge field [corresponding to longitudinal polarization] in quantum field theory.

Similarly from eqs. (\ref{total}) and (\ref{totalc}) we find that
\begin{eqnarray}
{\cal M}_{\rm non-eikonal}=\frac{k^\nu \gamma_\nu \gamma_\mu \epsilon^\mu(k)}{2q \cdot k} u(q)=\frac{k^\nu \gamma_\nu \gamma_\mu \epsilon_{\rm phys}^\mu(k)}{2q \cdot k} u(q) \rightarrow {\rm finite} ~~~~~~~~~~~{\rm as}~~~~~~~~~~(k_0,k_1,k_2,k_3) \rightarrow 0\nonumber \\
\label{totale}
\end{eqnarray}
which contribute to the finite part of the cross section which implies that pure gauge field [corresponding to longitudinal
polarization] does not contribute to the finite cross section and the finite cross section can be calculated by using
physical gauge field [corresponding to transverse polarization].

Hence we find that the non-eikonal-line part of the diagram as given by eq. (\ref{totale}) is necessary if we are calculating
the finite value of the cross section but it is not necessary if we are calculating the relevant infrared divergence behavior. The relevant
infrared divergence behavior can be calculated by using the eikonal-line part of the diagram as given by eq. (\ref{totald}).

For this reason, in the proof of NRQCD factorization of infrared divergences for heavy quarkonium production at NNLO in coupling constant,
the non-eikonal-line part of the diagram as given by eq. (\ref{totale}) is not considered as the full calculation of the cross section
or fragmentation function at NNLO will be daunting but fortunately the analysis of relevant infrared behavior at NNLO requires only the eikonal
approximation as given by eq. (\ref{totald}), see the discussion in the last paragraph of section 4 of \cite{nyka}.
Similarly the full calculation of the cross section or fragmentation
function at all order in coupling constant by using the non-eikonal-line part of the diagram as given by eq. (\ref{totale})
will require non-perturbative QCD which is not solved yet but fortunately the analysis of relevant infrared behavior at
at all order in coupling constant requires only the eikonal approximation as given by eq. (\ref{totald}).

Hence we find that we do not need to calculate the finite value of the cross section
(or the full cross section) [which will require the non-eikonal-line part of the diagram as given
by eq. (\ref{totale})] to study the relevant infrared divergence behavior. The relevant infrared
divergence behavior can be calculated by using eikonal approximation as given by eq. (\ref{totald}).

From eq. (\ref{totale}) we find that
\begin{eqnarray}
{\cal M}^{\rm pure~gauge~field}_{\rm non-eikonal}=\frac{k^\nu \gamma_\nu \gamma_\mu \epsilon^\mu_{\rm pure}(k)}{2q \cdot k} u(q)=0.
\label{totalf}
\end{eqnarray}
We are interested in the infrared divergence behavior due to the presence of the light-like Wilson line.
We have shown in eqs. (\ref{eeffl}) and (\ref{ffl}) that the eikonal current of the light-like charge generates pure gauge
field in quantum field theory. Hence from eqs. (\ref{eeffl}), (\ref{totald}), (\ref{ffl}) and (\ref{totalf}) we find that the
soft (infrared) divergence behavior due to the presence light-like Wilson line
can be studied by using pure gauge field in quantum field theory without modifying the finite value of the cross section.

\section{ Heavy quark-antiquark non-perturbative matrix element in the presence of Light-Like Wilson Line in QCD }

We have seen in section V that the infrared behavior of the non-perturbative NRQCD matrix element
$<0|\chi^\dagger K_n \xi (a^\dagger_H a_H) \xi^\dagger K_n' \chi|0>$ in eq. (\ref{bodwin})
can be obtained by studying the infrared behavior of the non-perturbative matrix element in QCD of the type
$<0|{\bar \Psi}(x) O_n \Psi(x) {\bar \Psi}(x') O'_n \Psi(x')|0>$
where $O_n,~O'_n$ are appropriate factors which identify the state of the
heavy quark-antiquark system such as the color singlet state or color octet state etc..
Similarly, we have also seen in section V that the infrared behavior of the non-perturbative matrix element in QCD of the type
$<0|{\bar \Psi}(x) O_n \Psi(x) {\bar \Psi}(x') O'_n \Psi(x')|0>$
due to the presence of light-like Wilson line in QCD can be studied by using the path integral method
of the QCD in the presence of SU(3) pure gauge background field. Hence we use the
path integral formulation of the background field method of QCD to study non-perturbative matrix element
$<0|{\bar \Psi}(x) O_n \Psi(x) {\bar \Psi}(x') O'_n \Psi(x')|0>$ in QCD in the presence of SU(3) pure gauge background
field as given by eq. (\ref{gtqcd}).

Background field method of QCD was originally formulated by 't Hooft \cite{thooft} and later
extended by Klueberg-Stern and Zuber \cite{zuber,zuber1} and by Abbott \cite{abbott}.
This is an elegant formalism which can be useful to construct gauge invariant
non-perturbative green's functions in QCD. This formalism is also useful to study quark and gluon production from classical chromo field \cite{peter}
via Schwinger mechanism \cite{schw}, to compute $\beta$ function in QCD \cite{peskin}, to perform
calculations in lattice gauge theories \cite{lattice} and to study evolution of QCD
coupling constant in the presence of chromofield \cite{nayak}.

In the background field method of QCD the generating functional is given by \cite{thooft,zuber,abbott}
\bea
&& Z[A,J,\eta,{\bar \eta}]=\int [dQ] [d{\bar \psi}] [d \psi ] ~{\rm det}(\frac{\delta G^a(Q)}{\delta \omega^b}) \nonumber \\
&& e^{i\int d^4x [-\frac{1}{4}{F^a}_{\mu \nu}^2[A+Q] -\frac{1}{2 \alpha}
(G^a(Q))^2+{\bar \psi} [i\gamma^\mu \partial_\mu -m +gT^a\gamma^\mu (A+Q)^a_\mu] \psi + J \cdot Q +{\bar \eta} \psi +  {\bar \psi} \eta]}
\label{zaqcd}
\eea
where $Q^{\mu a}(x)$ is the quantum gluon field and the gauge fixing term is given by
\bea
G^a(Q) =\partial_\mu Q^{\mu a} + gf^{abc} A_\mu^b Q^{\mu c}=D_\mu[A]Q^{\mu a}
\label{ga}
\eea
which depends on the background field $A^{\mu a}$ and
\bea
F_{\mu \nu}^a[A+Q]=\partial_\mu [A_\nu^a+Q_\nu^a]-\partial_\nu [A_\mu^a+Q_\mu^a]+gf^{abc} [A_\mu^b+Q_\mu^b][A_\nu^c+Q_\nu^c].
\eea
We have followed the notations of \cite{thooft,zuber,abbott} and accordingly we have
denoted the quantum gluon field by $Q^{\mu a}$ and the background field by $A^{\mu a}$.
The determinant ${\rm det}(\frac{\delta G^a(Q)}{\delta \omega^b})$ in eq. (\ref{zaqcd}) can be
expressed in terms of path integration over the ghost fields \cite{muta,zuber}. However, we will directly work with the
determinant ${\rm det}(\frac{\delta G^a(Q)}{\delta \omega^b})$ in eq. (\ref{zaqcd}).

Note that the gauge fixing term $\frac{1}{2 \alpha} (G^a(Q))^2$ in eq. (\ref{zaqcd}) [where $G^a(Q)$ is given by eq. (\ref{ga})]
is invariant for gauge transformation of $A_\mu^a$:
\bea
\delta A_\mu^a = gf^{abc}A_\mu^b\omega^c + \partial_\mu \omega^a,  ~~~~~~~({\rm type~ I ~transformation})
\label{typeI}
\eea
provided one also performs a homogeneous transformation of $Q_\mu^a$ \cite{zuber,abbott}:
\bea
\delta Q_\mu^a =gf^{abc}Q_\mu^b\omega^c.
\label{omega}
\eea
The gauge transformation of background field $A_\mu^a$ as given by eq. (\ref{typeI})
along with the homogeneous transformation of $Q_\mu^a$ in eq. (\ref{omega}) gives
\bea
\delta (A_\mu^a+Q_\mu^a) = gf^{abc}(A_\mu^b+Q_\mu^b)\omega^c + \partial_\mu \omega^a
\label{omegavbxn}
\eea
which leaves $-\frac{1}{4}{F^a}_{\mu \nu}^2[A+Q]$ invariant in eq. (\ref{zaqcd}).

For fixed $A_\mu^a$, {\it i.e.}, for
\bea
&&\delta A_\mu^a =0,  ~~~~~~~({\rm type~ II ~transformation})
\label{typeII}
\eea
the gauge transformation of $Q_\mu^a$ \cite{zuber,abbott}:
\bea
&&\delta Q_\mu^a = gf^{abc}(A_\mu^b + Q_\mu^b)\omega^c + \partial_\mu \omega^a
\label{omegaII}
\eea
gives eq. (\ref{omegavbxn}) which leaves $-\frac{1}{4}{F^a}_{\mu \nu}^2[A+Q]$ invariant in eq. (\ref{zaqcd}).

Extending eq. (\ref{zaqcd}) to include heavy quark [by using the lagrangian density from eq. (\ref{fullqcd})] we find that the
generating functional in the background field method of QCD is given by
\bea
&& Z[A,J,\eta_u,{\bar \eta}_u,\eta_d,{\bar \eta}_d,\eta_s,{\bar \eta}_s,\eta_h, {\bar \eta}_h]=\int [dQ] [d{\bar \psi}_1] [d \psi_1 ] [d{\bar \psi}_2] [d \psi_2 ][d{\bar \psi}_3] [d \psi_3 ][d{\bar \Psi}] [d \Psi ]~{\rm det}(\frac{\delta G^a(Q)}{\delta \omega^b}) \nonumber \\
&& {\rm exp}[i\int d^4x [-\frac{1}{4}{F^a}_{\mu \nu}^2[A+Q] -\frac{1}{2 \alpha}
(G^a(Q))^2+ J \cdot Q \nonumber \\
&&+\sum_{l=1}^3\left[{\bar \psi}_l [i\gamma^\mu \partial_\mu -m_l +gT^a\gamma^\mu (A+Q)^a_\mu] \psi_l +{\bar \eta}_l \psi_l +  {\bar \psi}_l\eta_l\right] \nonumber \\
&&+{\bar \Psi} [i\gamma^\mu \partial_\mu -M +gT^a\gamma^\mu (A+Q)^a_\mu] \Psi+{\bar \eta}_h \Psi + {\bar \Psi}\eta_h]].
\label{azaqcd}
\eea
Note that in the absence of external sources a pure gauge can be gauged away from
the generating functional. However, in the presence of external sources a pure gauge can not be gauged
away from the generating functional. It is useful to remember that, unlike QED \cite{tucci}, finding an exact relation between the generating
functional $Z[J,\eta_u,{\bar \eta}_u,\eta_d,{\bar \eta}_d,\eta_s,{\bar \eta}_s,\eta_h, \eta_h]$ in QCD in eq. (\ref{aqcd}) and the generating functional $Z[A,J,\eta_u,{\bar \eta}_u,\eta_d,{\bar \eta}_d,\eta_s,{\bar \eta}_s,\eta_h, \eta_h]$ in the background field method of QCD in eq. (\ref{azaqcd}) in the presence of SU(3) pure gauge background field is not easy.
The main difficulty is due to the gauge fixing terms which are different in both the cases. While the Lorentz (covariant) gauge fixing
term  $-\frac{1}{2 \alpha}(\partial_\mu Q^{\mu a})^2$ in eq. (\ref{aqcd}) in QCD is independent of the background field
$A^{\mu a}(x)$, the background field gauge fixing term $-\frac{1}{2 \alpha}(G^a(Q))^2$ in eq. (\ref{azaqcd}) in the background field method
of QCD depends on the background field $A^{\mu a}(x)$ where $G^a(Q)$ is given by eq. (\ref{ga}) \cite{thooft,zuber,abbott}.
Hence in order to study non-perturbative matrix element
in the background field method of QCD in the presence of SU(3) pure gauge background
field we proceed as follows.

By changing $Q \rightarrow Q-A$ in eq. (\ref{azaqcd}) we find that
\bea
&& Z[A,J,\eta_u,{\bar \eta}_u,\eta_d,{\bar \eta}_d,\eta_s,{\bar \eta}_s,\eta_h, {\bar \eta}_h]\nonumber \\
&&=e^{-i \int d^4x J \cdot A }\int [dQ] [d{\bar \psi}_1] [d \psi_1 ] [d{\bar \psi}_2] [d \psi_2 ][d{\bar \psi}_3] [d \psi_3 ][d{\bar \Psi}] [d \Psi ] ~{\rm det}(\frac{\delta G^a_f(Q)}{\delta \omega^b}) ~ \nonumber \\
&&e^{i\int d^4x [-\frac{1}{4}{F^a}_{\mu \nu}^2[Q] -\frac{1}{2 \alpha}
(G^a_f(Q))^2+ J \cdot Q +\sum_{l=1}^3\left[{\bar \psi}_l [i\gamma^\mu \partial_\mu -m_l +gT^a\gamma^\mu Q^a_\mu] \psi_l +{\bar \eta}_l \psi_l + \eta_l {\bar \psi}_l\right] +{\bar \Psi} [i\gamma^\mu \partial_\mu -M +gT^a\gamma^\mu Q^a_\mu] \Psi+{\bar \eta}_h \Psi + {\bar \Psi}\eta_h]}\nonumber \\
\label{zaqcd1}
\eea
where the gauge fixing term from eq. (\ref{ga}) becomes
\bea
G_f^a(Q) =\partial_\mu Q^{\mu a} + gf^{abc} A_\mu^b Q^{\mu c} - \partial_\mu A^{\mu a}=D_\mu[A] Q^{\mu a} - \partial_\mu A^{\mu a},
\label{gfa}
\eea
and eq. (\ref{omega}) [by using eq. (\ref{typeI}), type I transformation \cite{zuber,abbott}] becomes
\bea
&&\delta Q^a_\mu = gf^{abc}Q_\mu^b\omega^c+ \partial_\mu \omega^a.
\label{theta}
\eea
The eqs. (\ref{gfa}) and (\ref{theta}) can also be derived by using type II transformation which can be seen as follows.
By changing $Q \rightarrow Q-A$ in eq. (\ref{azaqcd}) we find eq. (\ref{zaqcd1})
where the gauge fixing term from eq. (\ref{ga}) becomes eq. (\ref{gfa})
and eq. (\ref{omegaII}) [by using eq. (\ref{typeII})] becomes eq. (\ref{theta}).
Hence we obtain eqs. (\ref{zaqcd1}), (\ref{gfa}) and (\ref{theta}) whether we use
the type I transformation or type II transformation. Hence we find that we will obtain the same
eq. (\ref{zaqcd1cz3}) whether we use the type I transformation or type II transformation.

Note that
\bea
A'^a_\mu(x) = A^a_\mu(x) +gf^{abc}\omega^c(x) A_\mu^b(x) + \partial_\mu \omega^a(x)
\label{athetaav}
\eea
in eq. (\ref{typeI}) is valid for infinitesimal transformation ($\omega << 1$) which is obtained from the
finite equation
\bea
T^aA'^a_\mu(x) = U(x)T^aA^a_\mu(x) U^{-1}(x)+\frac{1}{ig}[\partial_\mu U(x)] U^{-1}(x),~~~~~~~~~~~U(x)=e^{igT^a\omega^a(x)}.
\label{aftgrm}
\eea
Simplifying infinite numbers of non-commuting terms we find
\bea
\left[~e^{-igT^b\omega^b(x)} ~T^a~ e^{igT^c \omega^c(x)}~\right]_{ij}=[e^{-gM(x)}]_{ab}T^b_{ij}
\label{non}
\eea
where
\bea
M_{ab}(x)=f^{abc}\omega^c(x).
\label{mab}
\eea
Hence from eqs. (\ref{aftgrm}), (\ref{non}) and \cite{nayakj} we find that
\bea
{A'}_\mu^a(x) = [e^{gM(x)}]_{ab}A_\mu^b(x) ~+ ~[\frac{e^{gM(x)}-1}{gM(x)}]_{ab}~[\partial_\mu \omega^b(x)]
\label{ate}
\eea
where $M_{ab}(x)$ is given by eq. (\ref{mab}).
Similarly, the equation
\bea
Q'^a_\mu(x)= Q^a_\mu(x) +gf^{abc}\omega^c(x) Q_\mu^b(x) + \partial_\mu \omega^a(x)
\label{thetaav}
\eea
in eq. (\ref{theta}) is valid for infinitesimal transformation ($\omega << 1$) which is obtained from the
finite equation
\bea
T^aQ'^a_\mu(x) = U(x)T^aQ^a_\mu(x) U^{-1}(x)+\frac{1}{ig}[\partial_\mu U(x)] U^{-1}(x)
\label{ftgrm}
\eea
which gives
\bea
{Q'}_\mu^a(x) = [e^{gM(x)}]_{ab}Q_\mu^b(x) ~+ ~[\frac{e^{gM(x)}-1}{gM(x)}]_{ab}~[\partial_\mu \omega^b(x)]
\label{te}
\eea
where $M_{ab}(x)$ is given by eq. (\ref{mab}).

Changing the variables of integration from unprimed to primed variables in eq. (\ref{zaqcd1}) we find
\bea
&& Z[A,J,\eta_u,{\bar \eta}_u,\eta_d,{\bar \eta}_d,\eta_s,{\bar \eta}_s,\eta_h, {\bar \eta}_h]\nonumber \\
&&=e^{-i \int d^4x J \cdot A }\int [dQ'] [d{\bar \psi}'_1] [d \psi'_1 ] [d{\bar \psi}'_2] [d \psi'_2 ][d{\bar \psi}'_3] [d \psi'_3 ][d{\bar \Psi}'] [d \Psi' ] ~{\rm det}(\frac{\delta G^a_f(Q')}{\delta \omega^b}) ~ \nonumber \\
&&e^{i\int d^4x [-\frac{1}{4}{F^a}_{\mu \nu}^2[Q'] -\frac{1}{2 \alpha}
(G^a_f(Q'))^2+ J \cdot Q' +\sum_{l=1}^3\left[{\bar \psi}'_l [i\gamma^\mu \partial_\mu -m_l +gT^a\gamma^\mu Q'^a_\mu] \psi'_l +{\bar \eta}_l \psi'_l +{\bar \psi}'_l \eta_l \right] +{\bar \Psi}' [i\gamma^\mu \partial_\mu -M +gT^a\gamma^\mu Q'^a_\mu] \Psi'+{\bar \eta}_h \Psi' + {\bar \Psi}'\eta_h]}\nonumber \\
&&.
\label{zaqcd1b}
\eea
This is because a change of variables from unprimed to primed variables does not change the value of the
integration.

Under the finite transformation, using eq. (\ref{te}), we find
\bea
&& [dQ'] =[dQ] ~{\rm det} [\frac{\partial {Q'}^a}{\partial Q^b}] = [dQ] ~{\rm det} [[e^{gM(x)}]]=[dQ] {\rm exp}[{\rm Tr}({\rm ln}[e^{gM(x)}])]=[dQ]
\label{dqa}
\eea
where we have used (for any matrix $H$)
\bea
{\rm det}H={\rm exp}[{\rm Tr}({\rm ln}H)].
\eea

Similarly the fermion fields transform accordingly, see eq. (\ref{phg3}), {\it i.e.},
\bea
\psi'_l(x)=e^{igT^a\omega^a(x)}\psi_l(x),~~~~~~~~~~~~~\Psi'(x)=e^{igT^a\omega^a(x)}\Psi(x).
\label{pg3}
\eea
Using eqs. (\ref{te}) and (\ref{pg3}) we find
\bea
&&[d{\bar \psi}_1'] [d \psi'_1 ]=[d{\bar \psi}_1] [d \psi_1 ],~~~~~~~[d{\bar \psi}'_2] [d \psi'_2 ]=[d{\bar \psi}_2] [d \psi_2 ],~~~~~~[d{\bar \psi}'_3] [d \psi'_3 ]=[d{\bar \psi}_3] [d \psi_3 ],\nonumber \\
&&[d{\bar \Psi}'] [d \Psi' ]=[d{\bar \Psi}] [d \Psi ],~~~~~~{\bar \psi}'_l [i\gamma^\mu \partial_\mu -m_l +gT^a\gamma^\mu Q'^a_\mu] \psi'_l={\bar \psi}_l [i\gamma^\mu \partial_\mu -m_l +gT^a\gamma^\mu Q^a_\mu] \psi_l, \nonumber \\
&&{\bar \Psi}' [i\gamma^\mu \partial_\mu -M +gT^a\gamma^\mu Q'^a_\mu] \Psi'={\bar \Psi} [i\gamma^\mu \partial_\mu -m_l +gT^a\gamma^\mu Q^a_\mu]\Psi,~~~~~~~~~{F^a}_{\mu \nu}^2[Q']={F^a}_{\mu \nu}^2[Q].\nonumber \\
\label{psa}
\eea

Using eqs. (\ref{dqa}) and (\ref{psa}) in eq. (\ref{zaqcd1b}) we find
\bea
&& Z[A,J,\eta_u,{\bar \eta}_u,\eta_d,{\bar \eta}_d,\eta_s,{\bar \eta}_s,\eta_h, {\bar \eta}_h]\nonumber \\
&&=e^{-i \int d^4x J \cdot A }\int [dQ] [d{\bar \psi}_1] [d \psi_1 ] [d{\bar \psi}_2] [d \psi_2 ][d{\bar \psi}_3] [d \psi_3 ][d{\bar \Psi}] [d \Psi ]~{\rm det}(\frac{\delta G^a_f(Q')}{\delta \omega^b}) ~\nonumber \\
&& e^{i\int d^4x [-\frac{1}{4}{F^a}_{\mu \nu}^2[Q] -\frac{1}{2 \alpha}
(G^a_f(Q'))^2+ J \cdot Q' +\sum_{l=1}^3\left[{\bar \psi}_l [i\gamma^\mu \partial_\mu -m_l +gT^a\gamma^\mu Q^a_\mu] \psi_l +{\bar \eta}_l \psi'_l + {\bar \psi}'_l \eta_l \right] +{\bar \Psi} [i\gamma^\mu \partial_\mu -M +gT^a\gamma^\mu Q^a_\mu] \Psi+{\bar \eta}_h \Psi' + {\bar \Psi}'\eta_h]}.\nonumber \\
\label{zaqcd1cv}
\eea
From eq. (\ref{gfa}) we find
\bea
G_f^a(Q') =\partial_\mu Q^{' \mu a} + gf^{abc} A_\mu^b Q^{' \mu c} - \partial_\mu A^{\mu a}.
\label{gfap}
\eea
By simplifying the infinite number of non-commuting terms in the SU(3) pure gauge in eq. (\ref{gtqcd}) we find \cite{nayakj}
\bea
A^{\mu a}(x)=\partial^\mu \omega^b(x)\left[\frac{e^{gM(x)}-1}{gM(x)}\right]_{ab}
\label{pg4}
\eea
where $M_{ab}(x)$ is given by eq. (\ref{mab}).
By using eqs. (\ref{te}) and (\ref{pg4}) in eq. (\ref{gfap}) we find
\bea
&&G_f^a(Q') =\partial^\mu [[e^{gM(x)}]_{ab}Q_\mu^b(x) ~+ ~[\frac{e^{gM(x)}-1}{gM(x)}]_{ab}~[\partial_\mu \omega^b(x)]]\nonumber \\
&&+ gf^{abc} [\partial^\mu \omega^e(x)\left[\frac{e^{gM(x)}-1}{gM(x)}\right]_{be}] [[e^{gM(x)}]_{cd}Q_\mu^d(x) ~+ ~[\frac{e^{gM(x)}-1}{gM(x)}]_{cd}~[\partial_\mu \omega^d(x)]]\nonumber \\
&&- \partial_\mu [\partial^\mu \omega^b(x)\left[\frac{e^{gM(x)}-1}{gM(x)}\right]_{ab}]
\label{gfapa}
\eea
which gives
\bea
&&G_f^a(Q') =\partial^\mu [[e^{gM(x)}]_{ab}Q_\mu^b(x)]\nonumber \\
&&+ gf^{abc} [\partial^\mu \omega^e(x)\left[\frac{e^{gM(x)}-1}{gM(x)}\right]_{be}] [[e^{gM(x)}]_{cd}Q_\mu^d(x) ~+ ~[\frac{e^{gM(x)}-1}{gM(x)}]_{cd}~[\partial_\mu \omega^d(x)]].
\label{gfapb}
\eea
From eq. (\ref{gfapb}) we find
\bea
&&G_f^a(Q') =\partial^\mu [[e^{gM(x)}]_{ab}Q_\mu^b(x)]+ gf^{abc} [\partial^\mu \omega^e(x)\left[\frac{e^{gM(x)}-1}{gM(x)}\right]_{be}] [[e^{gM(x)}]_{cd}Q_\mu^d(x)]
\label{gfapc}
\eea
which gives
\bea
&&G_f^a(Q') = [e^{gM(x)}]_{ab}\partial^\mu Q_\mu^b(x)\nonumber \\
&&+Q_\mu^b(x)\partial^\mu [[e^{gM(x)}]_{ab}]+  [\partial^\mu \omega^e(x)\left[\frac{e^{gM(x)}-1}{gM(x)}\right]_{be}]gf^{abc} [[e^{gM(x)}]_{cd}Q_\mu^d(x)].
\label{gfapd}
\eea
From \cite{nayakj} we find
\bea
\partial^\mu [e^{igT^a\omega^a(x)}]_{ij}=ig[\partial^\mu \omega^b(x)]\left[\frac{e^{gM(x)}-1}{gM(x)}\right]_{ab}T^a_{ik}[e^{igT^c\omega^c(x)}]_{kj}
\label{pg4j}
\eea
which in the adjoint representation of SU(3) gives (by using $T^a_{bc}=-if^{abc}$)
\bea
[\partial^\mu e^{gM(x)}]_{ad}=[\partial^\mu \omega^e(x)]\left[\frac{e^{gM(x)}-1}{gM(x)}\right]_{be}gf^{bac}[e^{M(x)}]_{cd}
\label{pg4k}
\eea
where $M_{ab}(x)$ is given by eq. (\ref{mab}). Using eq. (\ref{pg4k}) in (\ref{gfapd}) we find
\bea
&&G_f^a(Q') = [e^{gM(x)}]_{ab}\partial^\mu Q_\mu^b(x)
\label{gfape}
\eea
which gives
\bea
(G_f^a(Q'))^2 = (\partial_\mu Q^{\mu a}(x))^2.
\label{gfapf}
\eea
Since for $n \times n$ matrices $A$ and $B$ we have
\bea
{\rm det}(AB)=({\rm det}A)({\rm det} B)
\eea
we find from eq. (\ref{gfape}) that
\bea
&&{\rm det} [\frac{\delta G_f^a(Q')}{\delta \omega^b}] ={\rm det}
[\frac{ \delta [[e^{gM(x)}]_{ac}\partial^\mu Q_\mu^c(x)]}{\delta \omega^b}]={\rm det}[
[e^{gM(x)}]_{ac}\frac{ \delta (\partial^\mu Q_\mu^c(x))}{\delta \omega^b}]\nonumber \\
&&=\left[{\rm det}[
[e^{gM(x)}]_{ac}]\right]~\left[{\rm det}[\frac{ \delta (\partial^\mu Q_\mu^c(x))}{\delta \omega^b}]\right]={\rm exp}[{\rm Tr}({\rm ln}[e^{gM(x)}])]~{\rm det}[\frac{ \delta (\partial_\mu Q^{\mu a}(x))}{\delta \omega^b}]\nonumber \\
&&={\rm det}[\frac{ \delta (\partial_\mu Q^{\mu a}(x))}{\delta \omega^b}].
\label{gqp4a}
\eea
Using eqs. (\ref{gfapf}) and (\ref{gqp4a}) in eq. (\ref{zaqcd1cv}) we find
\bea
&& Z[A,J,\eta_u,{\bar \eta}_u,\eta_d,{\bar \eta}_d,\eta_s,{\bar \eta}_s,\eta_h, {\bar \eta}_h]\nonumber \\
&&=e^{-i \int d^4x J \cdot A }\int [dQ] [d{\bar \psi}_1] [d \psi_1 ] [d{\bar \psi}_2] [d \psi_2 ][d{\bar \psi}_3] [d \psi_3 ][d{\bar \Psi}] [d \Psi ] ~{\rm det}[\frac{ \delta (\partial_\mu Q^{\mu a}(x))}{\delta \omega^b}] ~ \nonumber \\
&&e^{i\int d^4x [-\frac{1}{4}{F^a}_{\mu \nu}^2[Q] -\frac{1}{2 \alpha}
(\partial_\mu Q^{\mu a})^2+ J \cdot Q' +\sum_{l=1}^3\left[{\bar \psi}_l [i\gamma^\mu \partial_\mu -m_l +gT^a\gamma^\mu Q^a_\mu] \psi_l +{\bar \eta}_l \psi'_l + {\bar \psi}'_l \eta_l \right] +{\bar \Psi} [i\gamma^\mu \partial_\mu -M +gT^a\gamma^\mu Q^a_\mu] \Psi+{\bar \eta}_h \Psi' + {\bar \Psi}'\eta_h]}.\nonumber \\
\label{zaqcd1cz3}
\eea
From eqs. (\ref{pg4}) and (\ref{te}) we find
\bea
{Q'}_\mu^a(x) -A_\mu^a(x)= [e^{gM(x)}]_{ab}Q_\mu^b(x)
\label{tej}
\eea
where $M_{ab}(x)$ is given by eq. (\ref{mab}).

Note that eqs. (\ref{zaqcd1cz3}), (\ref{tej}) and (\ref{phg3}) are valid whether we use type I
transformation [see eqs. (\ref{typeI}) and (\ref{omega})] or type II transformation [see eqs. (\ref{typeII}) and (\ref{omegaII})].

However, since eq. (\ref{aftgrm}) is used to study the gauge transformation of the Wilson line in QCD, we will use
type I transformation [see eqs. (\ref{typeI}) and (\ref{omega})] in the rest of the paper which
for the finite transformation gives \cite{abbott,zuber}
\bea
J'^a_\mu(x)=[e^{gM(x)}]_{ab}J_\mu^b(x)
\label{jpr}
\eea
where $M_{ab}(x)$ is given by eq. (\ref{mab}).
From eqs. (\ref{zaqcd1cz3}), (\ref{tej}) and (\ref{jpr}) we find
\bea
&& Z[A,J',\eta_u,{\bar \eta}_u,\eta_d,{\bar \eta}_d,\eta_s,{\bar \eta}_s,\eta_h, {\bar \eta}_h]\nonumber \\
&&
~=~\int [dQ] [d{\bar \psi}_1] [d \psi_1 ] [d{\bar \psi}_2] [d \psi_2 ][d{\bar \psi}_3] [d \psi_3 ][d{\bar \Psi}] [d \Psi ] ~{\rm det}[\frac{ \delta (\partial_\mu Q^{\mu a}(x))}{\delta \omega^b}] ~ \nonumber \\
&&e^{i\int d^4x [-\frac{1}{4}{F^a}_{\mu \nu}^2[Q] -\frac{1}{2 \alpha}
(\partial_\mu Q^{\mu a})^2+ J \cdot Q +\sum_{l=1}^3\left[{\bar \psi}_l [i\gamma^\mu \partial_\mu -m_l +gT^a\gamma^\mu Q^a_\mu] \psi_l +{\bar \eta}_l \psi'_l + {\bar \psi}'_l \eta_l \right] +{\bar \Psi} [i\gamma^\mu \partial_\mu -M +gT^a\gamma^\mu Q^a_\mu] \Psi+{\bar \eta}_h \Psi' + {\bar \Psi}'\eta_h]}.\nonumber \\
\label{zaqcd1cz}
\eea

Under the non-abelian gauge transformation the fermion sources transform as \cite{abbott,zuber}
\bea
\eta'_l(x)=e^{igT^a\omega^a(x)}\eta_l(x),~~~~~~~~~~~~~\eta'_h(x)=e^{igT^a\omega^a(x)}\eta_h(x).
\label{pgg3}
\eea
From eqs. (\ref{pg3}) and (\ref{pgg3}) we find
\bea
{\bar \eta}'_l \psi'_l={\bar \eta}_l \psi_l,~~~~~~~~{\bar \psi}'_l\eta'_l={\bar \psi}_l\eta_l,~~~~~~~~~~{\bar \eta}'_h \Psi'={\bar \eta}_h \Psi,~~~~~~~~{\bar \Psi}'\eta'_h={\bar \Psi}\eta_h
\label{ppl}
\eea
which gives from eq. (\ref{zaqcd1cz})
\bea
&& Z[A,J',\eta'_u,{\bar \eta}'_u,\eta'_d,{\bar \eta}'_d,\eta'_s,{\bar \eta}'_s,\eta'_h, {\bar \eta}'_h]\nonumber \\
&&
~=~\int [dQ] [d{\bar \psi}_1] [d \psi_1 ] [d{\bar \psi}_2] [d \psi_2 ][d{\bar \psi}_3] [d \psi_3 ][d{\bar \Psi}] [d \Psi ] ~{\rm det}[\frac{ \delta (\partial_\mu Q^{\mu a}(x))}{\delta \omega^b}] ~ \nonumber \\
&&e^{i\int d^4x [-\frac{1}{4}{F^a}_{\mu \nu}^2[Q] -\frac{1}{2 \alpha}
(\partial_\mu Q^{\mu a})^2+ J \cdot Q +\sum_{l=1}^3\left[{\bar \psi}_l [i\gamma^\mu \partial_\mu -m_l +gT^a\gamma^\mu Q^a_\mu] \psi_l +{\bar \eta}_l \psi_l + {\bar \psi}_l \eta_l \right] +{\bar \Psi} [i\gamma^\mu \partial_\mu -M +gT^a\gamma^\mu Q^a_\mu] \Psi+{\bar \eta}_h \Psi + {\bar \Psi}\eta_h]}.\nonumber \\
\label{zaqcd1c}
\eea
Hence from eqs. (\ref{zaqcd1c}) and (\ref{aqcd}) we find
\bea
Z[J,\eta_u,{\bar \eta}_u,\eta_d,{\bar \eta}_d,\eta_s,{\bar \eta}_s,\eta_h, {\bar \eta}_h]=Z[A,J',\eta'_u,{\bar \eta}'_u,\eta'_d,{\bar \eta}'_d,\eta'_s,{\bar \eta}'_s,\eta'_h, {\bar \eta}'_h]
\label{final}
\eea
when the background field $A^{\mu a}(x)$ is the SU(3) pure gauge field as given by eq. (\ref{gtqcd}).

Hence we find that eq. (\ref{final}) is the
relation between the generating functional $Z[J,\eta_u,{\bar \eta}_u,\eta_d,{\bar \eta}_d,\eta_s,{\bar \eta}_s,\eta_h, {\bar \eta}_h]$
in QCD and the generating functional $Z[A,J,\eta_u,{\bar \eta}_u,\eta_d,{\bar \eta}_d,\eta_s,{\bar \eta}_s,\eta_h, {\bar \eta}_h]$
in the the background field method of QCD in the presence of SU(3) pure gauge background field $A^{\mu a}(x)$ as given by eq. (\ref{gtqcd}).

Note that in QED the corresponding result is \cite{tucci,nayakqed}
\bea
Z[J,\eta,{\bar \eta}]=Z[A,J,\eta',{\bar \eta}']
\label{finalqed}
\eea
when the background field $A^{\mu }(x)$ is the U(1) pure gauge field given by $A^\mu(x)=\partial^\mu \omega(x)$.
Eq. (\ref{tucci}) in QED is obtained from eq. (\ref{finalqed}).
Note that unlike eq. (\ref{final}) in QCD there is no $J'$ in eq. (\ref{finalqed}) in QED because while the
(quantum) gluon directly interacts with classical chromo-electromagnetic field the (quantum) photon does
not directly interact with classical electromagnetic field.

Eq. (\ref{final}) is the main result of this paper.

For the heavy quark Dirac field $\Psi(x)$,
the non-perturbative matrix element of the type $<0|{\bar \Psi}(x) O_n \Psi(x) {\bar \Psi}(x') O'_n \Psi(x')|0>$ in
QCD is given by eq. (\ref{alepg})
if the factors $O_n$ and $O'_n$ are independent of quantum fields. Similarly for the heavy quark Dirac field $\Psi(x)$,
the corresponding non-perturbative matrix element of the type $<0|{\bar \Psi}(x) O_n \Psi(x) {\bar \Psi}(x') O'_n \Psi(x')|0>$ in
the background field method of QCD is given by \cite{tucci}
\bea
&&<0|{\bar \Psi}(x) O_n \Psi(x) {\bar \Psi}(x') O'_n \Psi(x')|0>_A\nonumber \\
&&=\frac{\delta}{\delta \eta_h(x)}O_n\frac{\delta}{\delta {\bar \eta}_h(x)}\frac{\delta}{\delta \eta_h(x')}O'_n\frac{\delta}{\delta {\bar \eta}_h(x')}Z[A,J,\eta_u,{\bar \eta}_u,\eta_d,{\bar \eta}_d,\eta_s,\nonumber \\
&&{\bar \eta}_s,\eta_h, {\bar \eta}_h]|_{J=\eta_u={\bar \eta}_u=\eta_d={\bar \eta}_d=\eta_s={\bar \eta}_s=\eta_h= \eta_h=0}
\label{blepg}
\eea
where the suppression of the normalization factor $Z[0]$ is understood as it will cancel in the final result
(see eq. (\ref{finalfacts})).

When the background field $A^{\mu a}(x)$ is the SU(3) pure gauge as given by eq. (\ref{gtqcd})
we find from eqs. (\ref{alepg}), (\ref{blepg}), (\ref{final}), (\ref{pgg3}) and (\ref{jpr}) that
\bea
&&<0|{\bar \Psi}(x) O_n \Psi(x) {\bar \Psi}(x') O'_n \Psi(x')|0>\nonumber \\
&&=<0|{\bar \Psi}(x)\Phi(x) O_n\Phi^\dagger(x)\Psi(x) {\bar \Psi}(x')\Phi(x') O'_n \Phi^\dagger(x')\Psi(x')|0>_A
\label{finalfactsv}
\eea
if the factors $O_n$ and $O'_n$ are independent of quantum fields where, see eq. (\ref{ttt}),
\bea
\Phi(x)={\rm exp}[igT^a\omega^a(x)]={\cal P}e^{-ig \int_0^{\infty} d\lambda l\cdot { A}^a(x+l\lambda)T^a }.
\label{omng}
\eea
Note that the creation operator $a^\dagger_q$ and annihilation operator $a_q$ of the quark are related to the quark field
via the equation \cite{mandl}
\bea
&& \psi(x) = \sum_{\rm spin} \sum_p \sqrt{\frac{m}{VE_p}}
 [a_q(p) u(p) e^{-ip\cdot x} + a^\dagger_{\bar q}(p)
v(p) e^{ip\cdot x} ]
\eea
where color indices are suppressed. Hence one finds that the quark field $\psi(x)$ or $\Psi(x)$ depends on the
the creation (annihilation) operator $a^\dagger_q~(a_q)$ of the quark but is independent of the creation (annihilation)
operator $a^\dagger_H~(a_H)$ of the hadron. Similarly the gluon field $Q^{\mu a}(x)$ is independent of the creation
(annihilation) operator $a^\dagger_H~(a_H)$ of the hadron.
Since $a^\dagger_H a_H$ is independent of $\psi(x),\Psi(x),Q^{\mu a}(x)$
one can perform exactly the similar steps of the path integral calculation
as above to find from eq. (\ref{finalfactsv}) that
\bea
&&<0|{\bar \Psi}(x) O_n \Psi(x)a^\dagger_H a_H {\bar \Psi}(x') O'_n \Psi(x')|0>\nonumber \\
&&=<0|{\bar \Psi}(x)\Phi(x) O_n\Phi^\dagger(x)\Psi(x)a^\dagger_Ha_H {\bar \Psi}(x')\Phi(x') O'_n \Phi^\dagger(x')\Psi(x')|0>_A
\label{finalfacts}
\eea
where $\Phi(x)$ is given by eq. (\ref{omng}).

Under non-abelian gauge transformation as given by eq. (\ref{aftgrm}) the Wilson line in QCD transforms as
\bea
{\cal P}e^{ig \int_{x_i}^{x_f} dx^\mu A'^a_\mu(x)T^a }=U(x_f)\left[{\cal P}e^{ig \int_{x_i}^{x_f} dx^\mu A^a_\mu(x)T^a }\right]U^{-1}(x_i).
\label{ty}
\eea
From eqs. (\ref{lkjn}) and (\ref{ty}) we find
\bea
{\cal P}e^{-ig \int_0^{\infty} d\lambda l\cdot { A}'^a(x+l\lambda)T^a }=U(x){\cal P}e^{-ig \int_0^{\infty} d\lambda l\cdot { A}^a(x+l\lambda)T^a },~~~~~~~~~~~~~~U(x)={\rm exp}[igT^a\omega^a(x)]\nonumber \\
\label{kkkj}
\eea
which gives from eq. (\ref{omng})
\bea
\Phi'(x)=U(x)\Phi(x),~~~~~~~~~~~~~~~~~~~~~\Phi'^\dagger(x)=\Phi^\dagger(x)U^{-1}(x).
\label{hqg}
\eea
Hence we find that
$<0|{\bar \Psi}(x)\Phi(x) O_n\Phi^\dagger(x)\Psi(x)  a^\dagger_H a_H {\bar \Psi}(x')\Phi(x') O'_n \Phi^\dagger(x')\Psi(x')|0>_A$
in eq. (\ref{finalfacts})
is gauge invariant and eq. (\ref{finalfacts}) is consistent with the factorization of infrared divergences in QCD.

\section{ Proof of Factorization in heavy quarkonium production in NRQCD color octet mechanism at all order in Coupling constant }

The non-perturbative matrix element $<0|{\bar \Psi}(x) O_n \Psi(x) a^\dagger_H a_H {\bar \Psi}(x') O'_n \Psi(x')|0>$ in QCD in
eq. (\ref{finalfacts}) is obtained from the exact generating functional in QCD
as given by eq. (\ref{aqcd}), see eq. (\ref{alepg}). Similarly,
the non-perturbative matrix element $<0|{\bar \Psi}(x) O_n \Psi(x) a^\dagger_H a_H  {\bar \Psi}(x') O'_n \Psi(x')|0>_A$
in the background field method of QCD in eq. (\ref{finalfacts}) is obtained from the
exact generating functional in the background field method of QCD as given by
eq. (\ref{azaqcd}), see eq. (\ref{blepg}). Hence we find that eq. (\ref{finalfacts}) is valid at all order in coupling constant in QCD.

Note that in \cite{nyka,nykb} the proof of factorization is presented at NNLO in coupling constant and
to $v^2$ order in relative velocity of the heavy quark-antiquark pair. This is done by restricting the
result to $v^2$ order by using
\bea
p_1^\mu = \frac{P}{2} + p_r^\mu,~~~~~~~~~~p_2^\mu = \frac{P}{2} - p_r^\mu
\label{p12}
\eea
where $p^\mu_1$ is the momentum of the heavy quark, $p_2^\mu$ is the
momentum of the heavy antiquark, $P^\mu$ is the total momentum of the heavy quark-antiquark pair
and $p^\mu_r$ is the relative momentum of the heavy quark-antiquark pair. In the rest frame of
heavy quark-antiquark pair ${\vec p}_r=M{\vec v}$ where $M$ is the mass of the heavy quark.

Similarly in \cite{nayaksterman} the proof of factorization is presented at NNLO in coupling constant and
to all powers in relative velocity $v$ of the heavy quark-antiquark pair. This is done by obtaining the
result for arbitrary $p_1^\mu$ and $p_2^\mu$ without restricting to order $p_r^2$.
Hence in order to be consistent with the proof of factorization of \cite{nayaksterman} to all powers of relative velocity $v$
it is necessary to present the final result for arbitrary $p^\mu_1$ and $p^\mu_2$.

It can be seen that the non-perturbative matrix element $<0|{\bar \Psi}(x) O_n \Psi(x) a^\dagger_H a_H {\bar \Psi}(x') O'_n \Psi(x')|0>$ in QCD in
eq. (\ref{finalfacts}) is obtained from the exact generating functional in QCD
as given by eq. (\ref{aqcd}) without putting any restrictions on heavy quark and antiquark momenta, see eq. (\ref{alepg}). Similarly,
the non-perturbative matrix element $<0|{\bar \Psi}(x) O_n \Psi(x) a^\dagger_H a_H {\bar \Psi}(x') O'_n \Psi(x')|0>_A$
in the background field method of QCD in eq. (\ref{finalfacts}) is obtained from the
exact generating functional in the background field method of QCD as given by
eq. (\ref{azaqcd}) without putting any restrictions on heavy quark and antiquark momenta, see eq. (\ref{blepg}).
Hence we find that eq. (\ref{finalfacts}) is valid for any arbitrary momenta $p_1^\mu$ and
$p_2^\mu$ of the heavy quark and antiquark respectively. This implies that eq. (\ref{finalfacts}) is valid
to all powers in heavy quark relative velocity.

Hence we find that eq. (\ref{finalfacts}) is valid at all order in coupling constant in QCD and to all powers in the
heavy quark relative velocity.

As mentioned earlier, in NRQCD an ultraviolet cutoff $\Lambda \sim M$ is introduced \cite{nrqcd}. Hence the
ultraviolet (UV) behavior of QCD and NRQCD differ. However, the
infrared (IR) behavior of QCD and NRQCD remains same \cite{stewart}. Hence the
infrared behavior of the non-perturbative NRQCD matrix element
$<0|\chi^\dagger K_n \xi (a^\dagger_H a_H) \xi^\dagger K_n' \chi|0>$ in eq. (\ref{bodwin})
can be obtained by studying the infrared behavior of the non-perturbative matrix element in QCD of the type
$<0|{\bar \Psi}(x) O_n \Psi(x)  a^\dagger_H a_H {\bar \Psi}(x') O'_n \Psi(x')|0>$
where $O_n,~O'_n$ are appropriate factors which identify the state of the
heavy quark-antiquark system such as the color singlet state or color octet state etc..

We are interested in the effect of exchange of soft-gluons between the
light-like Wilson line and the heavy quark (and/or antiquark) in NRQCD color octet mechanism \cite{nyka,nykb,nayaksterman}.
Hence for the color singlet S-wave non-perturbative matrix element we find from eq. (\ref{finalfacts}) that
\bea
&&<0|{\cal O}_n|0>= <\chi^\dagger(0) K_{n} \xi(0) (a^\dagger_H a_H) \xi^\dagger(0) K'_{n} \chi(0)|0>
\label{nr2finsing}
\eea
at all order in coupling constant which is consistent with eq. (\ref{bodwin}).

When the factors $O_n$, $O'_n$ contain the color matrix $T^a$ we find
by simplifying infinite numbers of non-commuting terms [see eq. (\ref{non})] that eqs.
(\ref{finalfacts}) and (\ref{omng}) give
\bea
&&<0|{\bar \Psi}(x) O_{n,a} \Psi(x)a^\dagger_H a_H {\bar \Psi}(x') O'_{n,a} \Psi(x')|0>\nonumber \\
&&=<0|{\bar \Psi}(x) O_{n,e}\Psi(x) \Phi^{(A)\dagger}_{eb}(x) a^\dagger_H a_H \Phi^{(A)}(x')_{ba} {\bar \Psi}(x') O'_{n,a} \Psi(x')|0>_A
\label{finalfact}
\eea
where
\bea
\Phi^{(A)}(x)={\cal P}e^{-ig \int_0^{\infty} d\lambda l\cdot { A}^a(x+l\lambda)T^{(A)a }},~~~~~~(T^{(A)c})_{ab}=-if^{abc}.
\label{adjfins}
\eea

Hence from eqs. (\ref{finalfact}) and (\ref{adjfins}) we find that
the gauge invariant octet S-wave non-perturbative NRQCD matrix element which
is consistent with factorization of infrared divergences at all order in coupling constant and to all
powers in the heavy quark relative velocity is given by
\bea
<0|{\cal O}_n|0> = <0|\chi^\dagger(0) K_{n,e} \xi(0) \Phi_l^{(A)\dagger}(0)_{eb}(a^\dagger_H a_H) \Phi_l^{(A)}(0)_{ba}\xi^\dagger(0) K'_{n,a} \chi(0)|0>
\label{nrqcdfactfin}
\eea
where
\bea
\Phi_l^{(A)}(0)={\cal P}{\rm exp}[-igT^{(A)c}\int_0^{\infty} d\lambda l\cdot { A}^c(l\lambda)],~~~~~~(T^{(A)c})_{ab}=-if^{abc}.
\label{adjfin}
\eea
Note that the non-perturbative matrix element $<0|{\bar \Psi}(x) O_n \Psi(x)  a^\dagger_H a_H {\bar \Psi}(x') O'_n \Psi(x')|0>$
in the left hand side of eq. (\ref{finalfacts}) is independent of $l^\mu$. Hence all the $l^\mu$
dependence in $\Phi(x)$ defined by eq. (\ref{omng}) in the non-perturbative matrix element
$<0|{\bar \Psi}(x)\Phi(x) O_n\Phi^\dagger(x)\Psi(x)  a^\dagger_H a_H {\bar \Psi}(x')\Phi(x') O'_n \Phi^\dagger(x')\Psi(x')|0>_A$
in the right hand side of eq. (\ref{finalfacts}) is canceled by the use of background field $A^{\mu a}(x)$ in the expectation
value of the non-perturbative matrix element $<0|{\bar \Psi}(x) O_n \Psi(x) a^\dagger_H a_H  {\bar \Psi}(x') O'_n \Psi(x')|0>_A$
as defined in eq. (\ref{blepg}) in the the background field method of QCD. This proves that
the long-distance behavior of the non-perturbative NRQCD matrix element $<0|\chi^\dagger(0) K_{n,e} \xi(0) \Phi_l^{(A)\dagger}(0)_{eb}(a^\dagger_H a_H) \Phi_l^{(A)}(0)_{ba}\xi^\dagger(0) K'_{n,a} \chi(0)|0>$ in eq.
(\ref{nrqcdfactfin}) is independent of the light-like vector
$l^\mu$ at all order in coupling constant and to all powers in heavy quark relative velocity.

To summarize this, we find that eq. (\ref{nrqcdfactfin}), which is found by using path integral method of QCD,
is valid at all order in coupling constant and to all powers in heavy quark relative velocity.
We have also shown that
the long-distance behavior of the non-perturbative NRQCD matrix element is independent of the light-like vector $l^\mu$ at all order
in coupling constant and to all powers in heavy quark relative velocity.
The eq. (\ref{nrqcdfact}), which is found by using diagrammatic method of QCD
at NNLO in coupling constant and to all powers in heavy quark relative velocity
shows that long-distance behavior of the non-perturbative NRQCD matrix element
is independent of the light-like vector $l^\mu$ at NNLO
in coupling constant and to all powers in heavy quark relative velocity.
This implies that the gauge invariance and the
factorization at all order in coupling constant require gauge-completed octet S-wave non-perturbative NRQCD
matrix element that was introduced previously to prove factorization at NNLO.

Hence we find that eq. (\ref{nrqcdfact}) is valid at all order in coupling constant and to all
powers in the heavy quark relative velocity.

\section{Factorization Theorem is a key ingredient in calculation of NRQCD Heavy Quarkonium production cross section }

As mentioned earlier the
definition of the NRQCD heavy quarkonium production matrix element from heavy quark-antiquark pair
is a non-perturbative quantity which can not be calculated by using perturbation
theory no matter how many orders of perturbation
theory is used. From this point of view the path integral formulation (as opposed to diagrammatic methods in perturbation theory)
is useful to study the properties of the NRQCD non-perturbative matrix element of heavy quarkonium production
at all order in coupling constant. As mentioned earlier the only path integral formulation to study factorization of
soft and collinear divergences at all order in coupling constant in quantum field theory available
is by R. Tucci \cite{tucci}. However, the calculation of R. Tucci \cite{tucci} was exact for QED but was not exact for QCD.
We have extended the exact path integral calculation of proof of factorization of R. Tucci in QED \cite{tucci} to proof
of factorization in QCD at all order in coupling constant in \cite{nayaka3} and to proof
of factorization in NRQCD heavy quarkonium production at all order in coupling constant in the previous section.

In this section we will show how the factorization theorem as given by eqs. (\ref{finalfactsv})
and (\ref{finalfact}) is actually a key ingredient in calculation of NRQCD heavy quarkonium production
by using eqs. (\ref{css}) and (\ref{css1}) where the NRQCD non-perturbative matrix
element of heavy quarkonium production in color octet mechanism is given by eq. (\ref{nrqcdfactfin}).

Let us prove how the eqs. (\ref{finalfactsv}) and (\ref{finalfact}) are key ingredients
to prove eqs. (\ref{css}) and (\ref{css1}) to calculate the NRQCD heavy quarkonium production in color octet mechanism.
Suppose we calculate the cross section of heavy quark-antiquark production in color octet state
in the presence of light-like quark (or gluon). Then from eq. (\ref{finalfactsv}) we find
\bea
&&<0|{\bar \Psi}(x) O_n \Psi(x) {\bar \Psi}(x') O'_n \Psi(x')|0>_A\nonumber \\
&&=<0|{\bar \Psi}(x)\Phi^\dagger(x) O_n\Phi(x)\Psi(x) {\bar \Psi}(x')\Phi^\dagger(x') O'_n \Phi(x')\Psi(x')|0>.
\label{finalzibnst}
\eea
Hence from eq. (\ref{finalzibnst}) we find that the $A$ dependence which arises due to the soft gluon exchanges with
light-like quark (or gluon) is factorized and only appears in the gauge-links $\Phi(x)$ in the right hand side where
$\Phi(x)$ is given by eq. (\ref{omng}). Eq. (\ref{finalzibnst}) implies that in the cross section
for $Q{\bar Q}$ production in color octet state at all order in coupling constant the infrared divergences due to the presence of
light-like quark (or gluon) are factorized only to the gauge links $\Phi(x)$.

Eq. (\ref{finalzibnst}) is the exact extension of eq. (1.6) of \cite{tucci} of the factorization in QED.

Hence from eq. (\ref{finalzibnst}) we find that the non-perturbative matrix element of NRQCD heavy quarkonium production
in color octet mechanism which cancels these infrared divergences and is consistent with the factorization theorem
is obtained from eq. (\ref{finalfact}) and is given by eq. (\ref{nrqcdfactfin}). This proves
that the factorization theorem as given by eqs. (\ref{finalfactsv})
and (\ref{finalfact}) is actually a key ingredient to prove eqs. (\ref{css}) and (\ref{css1}) to calculate the
NRQCD heavy quarkonium production cross section in color octet mechanism at all order in coupling constant.

\section{Conclusions}
Recently the proof of factorization in heavy quarkonium production in NRQCD color octet mechanism is given
at next-to-next-to-leading order (NNLO) in coupling constant by using diagrammatic method of QCD. In this paper
we have proved factorization in heavy quarkonium production in NRQCD color octet mechanism at all order in coupling constant
by using path integral method of QCD. Our proof
is valid to all powers in the heavy quark relative velocity. We have found that the gauge invariance and the factorization at all
order in coupling constant require gauge-completed non-perturbative NRQCD
matrix elements that were introduced previously to prove factorization at NNLO.

\acknowledgments

I thank George Sterman for useful discussions and suggestions.

\appendix

\section{Interaction Between Non-Eikonal Current and the Gauge Field Generated by Light-Like Eikonal Current in Quantum Field Theory }

From the non-eikonal part of the diagram in eq. (\ref{noneik}) we find
\bea
&& e\int \frac{d^4k}{(2\pi)^4} \frac{k^\nu \gamma_\nu \gamma_\mu A^\mu(k)}{2q \cdot k+i\epsilon}
= \int d^4x J_\mu(x) A^\mu(x)
\eea
where the non-eikonal current density $J^\mu(x)$ of the (light-like or non-light-like) charge $e$ of four-momentum $q^\mu$ is given by
\bea
J^\mu(x) =\frac{e}{2} \gamma_\nu \gamma^\mu \int_0^{\infty} d\lambda  \frac{\partial}{\partial x_\nu } \delta^{(4)}(x-q\lambda).
\label{nekcd}
\eea
Hence using eqs. (\ref{nekcd}) and (\ref{ekcd}) in eq. (\ref{wj}) we find that the interaction between the non-eikonal current
and the gauge field generated by the light-like eikonal current in quantum field theory gives the effective (interaction)
action
\bea
&&{S}^{int}_{eff}[J] = l_\mu \frac{e^2}{4}\int d^4x \gamma_\nu \gamma^\mu \int_0^{\infty} d\lambda \delta^{(4)}(x-q\lambda) \frac{\partial}{\partial x_\nu }\frac{1}{\partial^2} \int d\lambda' ~\delta^{(4)}({ x}-l\lambda') \nonumber \\
&&= l^2 \frac{e^2}{2}\int d^4x [\frac{l \cdot \partial [q \cdot (x-q\lambda_0)]}{[q \cdot (x-q\lambda_0)]^2}][\frac{1}{(l \cdot x)^3}]
\label{effc}
\eea
where $\lambda_0$ is the solution of the equation
\bea
(x-q\lambda_0)^\mu (x-q\lambda_0)_\mu=0.
\label{l0}
\eea
From eqs. (\ref{effc}) and (\ref{l0}) we find that the interaction between the non-light-like non-eikonal current
and the gauge field generated by the light-like eikonal current in quantum field theory gives the effective (interaction)
lagrangian density
\bea
{\cal L}_{eff}^{int}(x) = l^2 \frac{e^2}{2}\frac{(l \cdot q) (q \cdot x) -(l \cdot x) q^2 }{(l \cdot x)^3[(q \cdot x)^2 -q^2 x^2]^{\frac{3}{2}}}.
\label{wjf}
\eea
From eq. (\ref{wjf}) we find that the interaction between the light-like non-eikonal current
and the gauge field generated by the light-like eikonal current in quantum field theory gives the effective (interaction)
lagrangian density
\bea
{\cal L}_{eff}^{int}(x) = \frac{e^2}{2} \frac{l^2 (q \cdot l) }{(q \cdot x)^2(l \cdot x)^3}.
\label{wjfl}
\eea

\end{document}